\documentstyle[prl,aps,preprint]{revtex}

\tighten
\begin{document}
\draft
 
\renewcommand{\thepage}{\empty}

\preprint{
\noindent
\hfill
\begin{minipage}[t]{3in}
\begin{flushright}
LBNL-39579\\
UCB-PTH-96/51\\
hep-ph/9612333 \\
December 1996
\end{flushright}
\end{minipage}
}

\title{Determining $\bbox{\tan\beta}$ from the SUSY Higgs Sector \\
at Future $\bbox{e^+e^-}$ Colliders}

\author{Jonathan L. Feng $^{ab}$\thanks
{Research Fellow, Miller Institute for Basic Research in
Science.}
and Takeo Moroi $^a$}

\address{\vspace*{.15in}
$^a$ Theoretical Physics Group, 
E.~O.~Lawrence Berkeley National Laboratory \\
Berkeley, California 94720 \\
\vspace*{.15in}
$^b$ Department of Physics, University of California \\
Berkeley, California 94720
}

\maketitle

\begin{abstract}

We examine the prospects for determining $\tan\beta$ from heavy Higgs
scalar production in the minimal supersymmetric standard model at a
future $e^+e^-$ collider.  Our analysis is independent of assumptions
of parameter unification, and we consider general radiative
corrections in the Higgs sector.  Bounds are presented for $\sqrt{s} =
500$ GeV and 1 TeV, several Higgs masses, a variety of integrated
luminosities and $b$-tagging efficiencies, and in scenarios with and
without supersymmetric decays of the Higgs bosons. We find stringent
constraints for $3 \alt \tan\beta \alt 10$, and, for some scenarios,
also interesting bounds on high $\tan\beta$ through $tbH^{\pm}$
production.  These bounds imply that simple Yukawa unifications may be
confirmed or excluded. Implications for soft scalar mass determination
and top squark parameters are also discussed.

\end{abstract}


\newpage
\renewcommand{\thepage}{\arabic{page}}
\narrowtext

\section{Introduction}
\label{sec:Introduction}

Supersymmetry (SUSY) is currently a promising framework for
understanding the physics of electroweak symmetry breaking, and its
discovery at future collider experiments is an exciting possibility.
In addition to elucidating weak scale physics, however, the discovery
of SUSY may also shed light on the mechanism of SUSY breaking, and may
even provide our first glimpse of physics at the grand unified theory
(GUT) and Planck scales.  The program of extrapolating weak scale
measurements to such high scales will be an extremely challenging one,
and its success is certainly not guaranteed.  What is likely, however,
is that such a program will require a detailed understanding of the
properties of the weak scale supersymmetric particles, or, in other
words, a precise determination of the various weak scale SUSY
parameters.

Of the many SUSY parameters, $\tan\beta$, the ratio of Higgs vacuum
expectation values, is important for a number of reasons. A
measurement of $\tan\beta$ allows one to determine Yukawa couplings,
and thereby confirm or exclude the possibilities of $m_b$--$m_\tau$
unification and SO(10)-like Yukawa unification \cite{b-tau,so(10)}.
The parameter $\tan\beta$ is also required to determine soft SUSY
breaking scalar masses from the measured physical sfermion masses.
Detailed knowledge of soft mass parameters may allow us to distinguish
various SUSY breaking mechanisms\cite{MEP}.  Finally, because the
parameter $\tan\beta$ enters all (neutralino/chargino, sfermion, and
scalar Higgs) sectors of SUSY theories, a precise measurement of
$\tan\beta$ from one sector allows one to check SUSY relations in
other sectors and improves the bounds on many other parameters.

In this paper we consider the prospects for determining $\tan\beta$
from the production of Higgs scalars in the general setting of a
supersymmetric model with minimal field content.  (For recent studies
in the framework of GUT scenarios, see Ref.~\cite{GK}.)  The branching
ratios of the heavy Higgs bosons are strongly dependent on $\tan\beta$
and may be weakly dependent on other SUSY parameters.  In contrast,
almost all other observables that depend on $\tan\beta$ also depend on
many additional unknown SUSY parameters, which weakens one's ability
to determine $\tan\beta$ precisely and in a model-independent way.  In
addition, heavy Higgs production results in an excess of multi-$b$
quark events, which, given the excellent $b$-tagging efficiency and
purity now expected to be available at future colliders, allows it to
be distinguished from standard model backgrounds.  This sector
therefore holds the promise of an exceptionally clean and powerful
determination of $\tan\beta$.  We will see that Higgs processes may
provide strong constraints on moderate and high $\tan\beta$, regions
which are particularly interesting for GUTs and soft scalar mass
determination.

In this study, we will consider the experimental setting of a future
$e^+e^-$ collider\cite{NLCPHYS,NLCZDR,JLCI}.  Such colliders have been
shown to be promising for Higgs discovery and
study\cite{JLCI,Komamiya,Gunionreview,MP}.  In particular, we will
consider two stages of the proposed Next Linear Collider (NLC): the
first with $\sqrt{s} = 500$ GeV and design luminosity $50 \text{
fb}^{-1}/\text{yr}$, and the second with $\sqrt{s} = 1$ TeV and
luminosity 100 -- 200 $\text{ fb}^{-1}/ \text{yr}$
\cite{NLCPHYS,NLCZDR}.  We will display results for a variety of
integrated luminosities, detector parameters, and systematic
uncertainties.

Our analysis is intended to estimate the power of particularly
promising processes for determining $\tan\beta$ in a general setting.
We do not restrict our attention to specific models by assuming
minimal supergravity boundary conditions or other parameter
unifications.  Rather, we analyze a number of scenarios by choosing
various kinematically accessible heavy Higgs masses and consider
scenarios with and without supersymmetric Higgs decays.  In addition,
we will discuss what improvements can be expected if information from
other experiments and processes are incorporated.  It should be
stressed, however, that if SUSY is discovered, the analysis should be
optimized for the particular SUSY parameters realized in nature, and
the SUSY parameters will best be determined by a global fit to all
data.

This paper is organized as follows.  In Sec.~\ref{sec:Tanb}, we
describe the dependence of Higgs processes on $\tan\beta$.  We explain
our general treatment of Higgs sector radiative corrections and
discuss the relevant cross sections and branching ratios.  Higgs
scalar production may be detected in a number of channels; we define
the channels we will use and the cuts used to isolate these signals in
Sec.~\ref{sec:Experiment}.  We then describe our Monte Carlo
simulations, and discuss the backgrounds and systematic errors
entering our analysis.  Sec.~\ref{sec:Results} contains our results
for $\tan\beta$ bounds in a variety of scenarios without
supersymmetric Higgs decays.  The effects of such decays are discussed
in Sec.~\ref{sec:SUSYdecays}.  Readers who are primarily interested in
our results are referred to the $(\tan\beta, \tan\beta')$ plots in
Secs.~\ref{sec:Results} and \ref{sec:SUSYdecays}.  Interesting
applications of these results are contained in Secs.~\ref{sec:Soft}
and \ref{sec:Upper}.  In Sec.~\ref{sec:Conclusions}, we briefly
compare our results to those that may be obtained with other processes
and present our conclusions.

\section{$\protect\bbox{\tan\beta}$ and the SUSY Higgs sector}
\label{sec:Tanb}

\subsection{Definitions and radiative corrections}
\label{subsec:Definitions}

We begin by reviewing the necessary details of the scalar Higgs
sector\cite{HHG}.  We consider a supersymmetric model with minimal
Higgs content, that is, with two Higgs superfields

\begin{equation}
 \hat {H}_1=\left ( \begin {array}{c} \hat {H}_1^0 \\
 \hat {H}_1^- \end {array} \right )
 \hspace {.2in} {\rm and} \hspace {.2in}
 \hat {H}_2=\left ( \begin {array}{c} \hat {H}_2^+ \\
 \hat {H}_2^0 \end {array} \right ) \ .
 \end{equation}
These couple through the superpotential

\begin{equation}
W = \lambda_E \hat{H}_1\hat{L}\hat{E}+\lambda_D \hat{H}_1\hat{Q}\hat{D}
    -\lambda_U \hat{H}_2\hat{Q}\hat{U} - \mu \hat{H}_1 \hat{H}_2 \ ,
\label{Superpot}
\end{equation}
where the $\lambda$ are Yukawa couplings and $\mu$ is the
supersymmetric Higgs mass parameter. The ratio of the two Higgs scalar
vacuum expectation values is

\begin{equation}
\tan \beta = \frac{ \langle H^0_2 \rangle }{ \langle H^0_1 \rangle } \ . 
\end{equation}
There are four physical Higgs scalars in the theory: the charged
scalar $H^{\pm}$, the CP--odd $A$, and the two CP--even scalars $h$
and $H$.  The CP--even mass matrix is in general given by

\begin{equation}
\label{massmatrix}
{\cal M}^2 = \left( \begin{array}{cc}
m_A^2 \sin^2\beta + m_Z^2 \cos^2\beta & - (m_A^2 + m_Z^2 ) 
\sin\beta \cos\beta \\
-(m_A^2 + m_Z^2 )\sin\beta\cos\beta & 
m_A^2 \cos^2\beta + m_Z^2 \sin^2\beta 
\end{array} \right) 
+ \Delta {\cal M}^2 \ ,
\end{equation}
in the basis $(H^0_1, H^0_2)$, where $\Delta {\cal M}^2$ contains all
the radiative corrections.  The mixing angle diagonalizing this
matrix, $\alpha$, enters in Higgs scalar pair production at $e^+e^-$
colliders through the vertices $ZAH$ and $ZZh$, which are proportional
to $\sin (\beta-\alpha)$, and the vertices $ZAh$ and $ZZH$, which are
proportional to $\cos (\beta-\alpha)$. In the limit of large $m_A$,
$\cos (\beta - \alpha) \to 0$.

At tree level, all Higgs scalar masses and interactions are completely
determined by $\tan\beta$ and one additional parameter, which is
conventionally taken to be the CP--odd mass $m_A$.  The charged Higgs
mass is then given by

\begin{equation}
\label{mhc=ma+mw}
m_{H^{\pm}}^2 = m_A^2 + m_W^2 \ ,
\end{equation}
and the CP--even masses and $\alpha$ are determined by
Eq.~(\ref{massmatrix}) with $\Delta {\cal M}^2 = 0$.  However, the
relations between Higgs masses and mixings receive radiative
corrections.  These corrections may be large in the CP--even
sector\cite{1loop}, and are dependent on many additional SUSY
parameters.  Precise measurements of other SUSY parameters, for
example, the masses and left-right mixing in the top squark sector,
may significantly constrain the size of these radiative corrections.
For most of this study, however, we make the conservative assumption
that no estimates of their size may be obtained from measurements
outside the Higgs sector, and we also do not assume that such effects
are small.

Given this framework, the Higgs masses and interactions are all
independent quantities, and must be determined experimentally.  In
this study, we will apply the results of previous analyses to
determine the Higgs masses\cite{Sop}, $\cos^2 (\beta- \alpha )$ from
$\sigma ( e^+e^- \to Zh)$\cite{JLCI}, and $B(H\to hh)$ from
$H\nu\bar{\nu}$ production\cite{DHZ}.  The uncertainties in these
measurements will be incorporated in our study as systematic errors.
Given these measurements, the only remaining unknown parameter
entering the processes we will study is $\tan\beta$.

By appealing to experimental measurements of Higgs masses and
interactions, we do not exploit theoretical relations between Higgs
masses and mixings to constrain $\tan\beta$, and we do not restrict
the applicability of our analysis to a specific set of parameters or
radiative corrections.  Ultimately, however, we must choose some
underlying parameters to study so that we may present quantitative
results.  The choice of parameters is guided by the desire to choose
scenarios that share qualitative features with a large portion of
parameter space.  In this study, we include the qualitative features
of radiative corrections by studying scenarios in which such effects
are given by the leading $m_t^4$ 1--loop contribution arising from a
top-stop loop without left-right stop mixing.  With this correction,
the radiative correction to the CP--even Higgs mass squared
matrix\cite{1loop} becomes

\begin{equation}
\label{deltamassmatrix}
\Delta {\cal M}^2 = \epsilon \left( \begin{array}{cc}
0 & 0 \\ 0 & m_Z^2 \end{array} \right) \ , 
\end{equation}
and the $Hhh$ vertex\cite{DHZ} becomes
\begin{equation}
\label{Hhh}
\lambda_{Hhh} = 2 \sin 2\alpha \sin (\beta+\alpha) - \cos 2 \alpha \cos
(\beta + \alpha) + 3\epsilon \frac{\sin\alpha}{\sin\beta}
\cos^2\alpha \ ,
\end{equation}
where 

\begin{equation}
\label{epsilon}
\epsilon = \frac{3}{8\pi^2}\frac{g^2}{\cos^2\theta_W}
\frac{1}{\sin^2\beta} \frac{m_t^4}{m_Z^4} \ln\left(
\frac{m_{\tilde{t}}^2}{m_t^2} \right) \ ,
\end{equation}
and $m_{\tilde{t}}= \sqrt{m_{\tilde{t}_1} m_{\tilde{t}_2}}$ is the
geometric mean of the two physical top squark masses.  The other
triple Higgs vertices also receive corrections, but these will not
enter our analysis. Furthermore, in this case, the tree level relation
between $m_{H^{\pm}}$ and $m_A$ given in Eq.~(\ref{mhc=ma+mw}) is not
affected. There are additional possible sources of (typically smaller)
radiative corrections, including the bottom squark sector, left-right
squark mixing, and the gaugino-Higgsino sector\cite{full1loop}.  Our
analysis procedure does not assume that these effects are absent, and
so could be applied to scenarios in which these effects are present as
well.  Of course, to the extent that these effects change the
underlying physics, the quantitative results presented in the
following sections will be modified.

Finally, aside from radiative corrections to the Higgs masses and
field compositions, there are corrections to the specific processes we
consider.  These are 1--loop corrections to the cross sections and
decay widths, which also may depend on unknown SUSY parameters, such
as squark masses.  We will include these effects as systematic errors,
and we discuss these errors more fully in Sec.~\ref{sec:Experiment}.

\subsection{Cross sections}

The two body production processes involving Higgs bosons at $e^+e^-$
colliders are $e^+e^- \to (\gamma^*, Z^*) \to H^+ H^-$ and $e^+e^- \to
Z^* \to Zh, ZH, Ah, AH$.  As noted in Sec.~\ref{subsec:Definitions},
production of $ZH$ and $Ah$ are suppressed by $\cos^2 (\beta-\alpha)$.
For $m_A \agt 200$ GeV, this is a large suppression, and, although
these processes have been included in our simulations, they are
statistically insignificant for this analysis.

In this study we will consider two energies for the NLC: $\sqrt{s} =
500$ GeV and 1 TeV.  We choose typical heavy Higgs masses within the
kinematically accessible range for each of these two energies. For the
500 GeV collider, we consider $m_{H^{\pm}}=200$ GeV, and for the 1 TeV
collider, we consider $m_{H^{\pm}}= 200$, 300, and 400 GeV.  (Here and
in the following, we choose to fix $m_{H^{\pm}}$ rather than the more
conventional $m_A$, as the charged Higgs will be seen to play the
central role in this analysis.  The CP--odd masses corresponding to
the choices above are $m_A = 183$, 289, and 392 GeV.)

\begin{figure}
\vspace*{.5in}\centerline{
\psfig{file=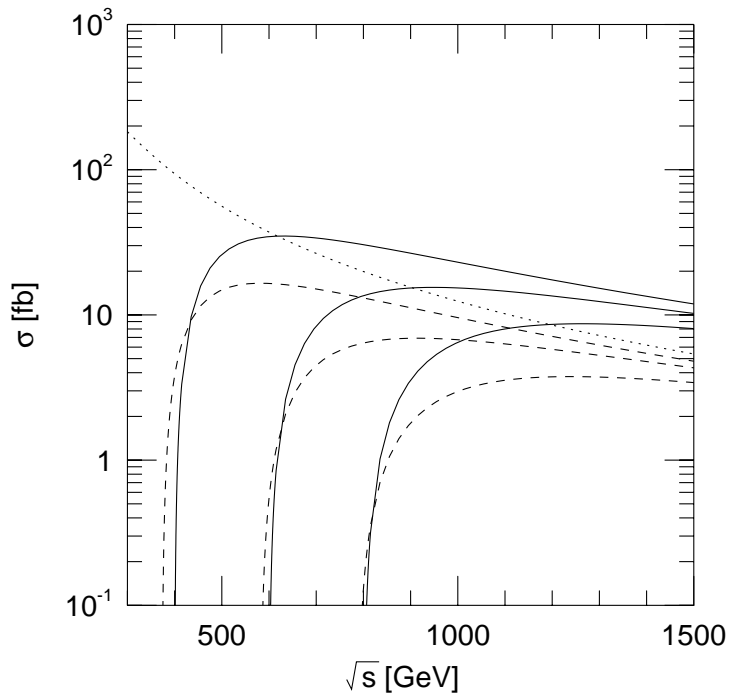,width=0.61\textwidth}
\vspace*{.5in}}
\caption{ Production cross sections for $e^+e^- \to H^+H^-$ (solid)
and $AH$ (dashed) for $m_{H^{\pm}} = 200$, 300, and 400 GeV from left
to right, and $Zh$ (dotted) for $m_{H^{\pm}} = 200$ GeV. (The $Zh$
cross sections for $m_{H^{\pm}} = 300$ and 400 GeV are virtually
identical.)  The 1--loop radiative correction given in
Eqs.~(\protect\ref{deltamassmatrix}) and (\protect\ref{epsilon}) has
been included with $m_{\tilde{t}}= 1$ TeV, and we have set $\tan\beta
= 5$; the dependence on $\tan\beta$ is very weak for $m_{H^{\pm}}
\protect\agt 200$ GeV.
\label{fig:production}}
\end{figure}

In Fig.~\ref{fig:production}, we plot the cross sections for $H^+H^-$,
$AH$, and $Zh$ as functions of the center of mass energy $\sqrt{s}$
for the three values of $m_{H^{\pm}}$ given above.  We have set
$\tan\beta = 5$; the dependence on $\tan\beta$ is very weak for
$m_{H^{\pm}} \agt 200$ GeV. For fixed $\tan\beta$ and $m_{H^{\pm}}$,
the Higgs masses and couplings are determined by including the 1--loop
radiative correction given in Eqs.~(\ref{deltamassmatrix}) and
(\ref{epsilon}) with $m_{\tilde{t}}= 1$ TeV.  We see that given the
NLC design luminosities of $50\text{ fb}^{-1} / \text{yr}$ and 100 --
200 $\text{ fb}^{-1} / \text{yr}$ for the two beam energies, and
assuming heavy Higgs masses sufficiently below threshold, thousands of
events per year will be produced through these reactions.

\begin{figure}
\vspace*{.5in}\centerline{
\psfig{file=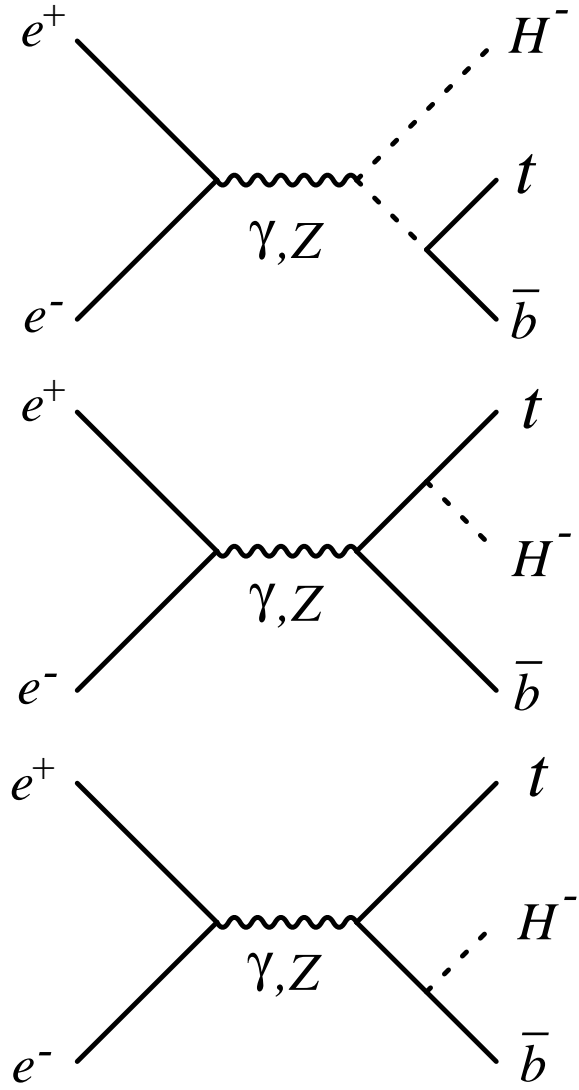,width=0.55\textwidth}
\vspace*{.5in}}
\caption{ The three Feynman diagrams contributing to the three body
final state $t\bar{b}H^-$.  Three similar diagrams contribute to
$\bar{t}b H^+$.
\label{fig:tbh}}
\end{figure}

\begin{figure}
\vspace*{.5in}\centerline{
\psfig{file=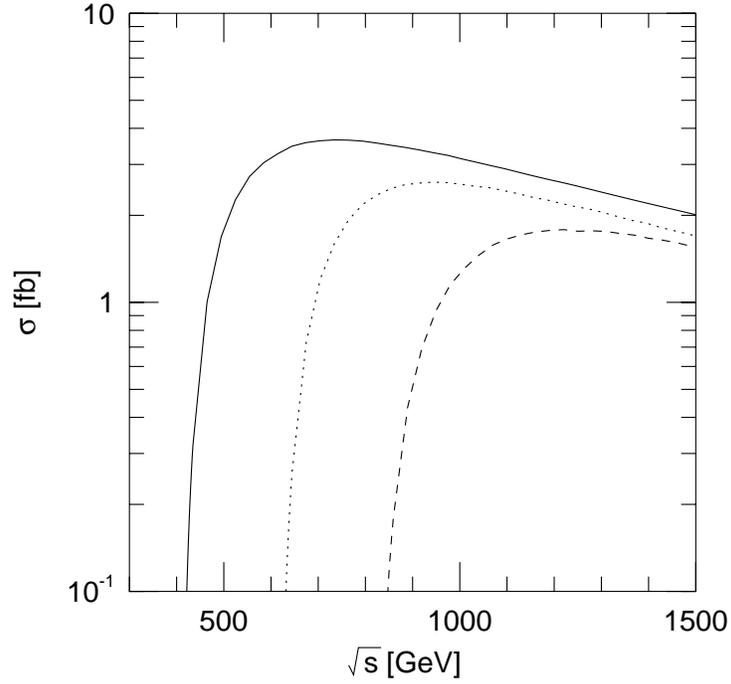,width=0.61\textwidth}
\vspace*{.5in}}
\caption{ The three-body cross section $\sigma(e^+e^- \to t\bar{b}H^-)
+ \sigma(e^+e^- \to \bar{t} b H^+)$ for $m_{H^{\pm}} = 200$ GeV
(solid), 300 GeV (dotted), and 400 GeV (dashed), with $\tan\beta=60$.
We require $E_t + E_b > 1.02 \protect\sqrt{s}/2$ to separate this mode
from the two-body production of $H^+H^-$ followed by $H^{\pm} \to tb$.
\label{fig:production2}}
\end{figure}

We will also make use of the three-body processes $e^+e^- \to
t\bar{b}H^- , \bar{t}b H^+$\cite{DKZ}.  This process takes place
through the Feynman diagrams of Fig.~\ref{fig:tbh}, and is greatly
enhanced through $m_b \tan\beta$ couplings for large $\tan\beta$.  The
production cross section is plotted in Fig.~\ref{fig:production2},
where we have set $\tan\beta=60$.  (In calculating the cross sections
for this figure, we have required $E_t + E_b > 1.02 \sqrt{s}/2$ to
separate this mode from the two-body production of $H^+H^-$ followed
by $H^{\pm} \to tb$; the detailed cuts we use in our analysis and
experimental simulation will be presented below.)  We will see that
for large $\tan\beta$, this mode may be extracted from backgrounds.
Its sensitivity to large $\tan\beta$ will then be useful for placing
$\tan\beta$ constraints in this range.  Although we will not study
them here, we note that $b\bar{b}A$ and $b\bar{b}H$ are similarly
enhanced for large $\tan\beta$, and may also be useful if they can be
isolated from backgrounds.

\subsection{Branching ratios}

\begin{figure}
\vspace*{.5in}\centerline{
\psfig{file=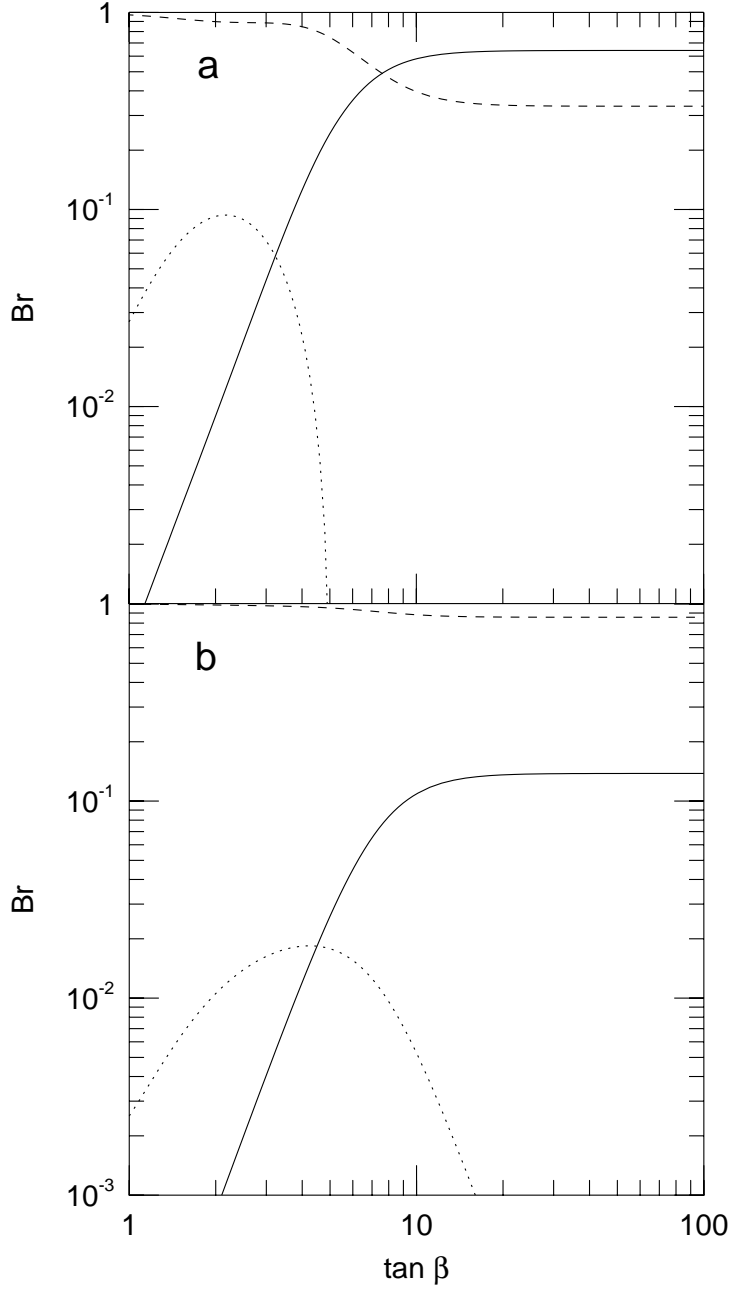,width=0.61\textwidth}
\vspace*{.5in}}
\caption{ The branching ratios for $H^-$ to $\tau \bar{\nu}$ (solid),
$\bar{t} b$ (dashed), and $W^- h$ (dotted) for (a) $m_{H^{\pm}}=200$
GeV and (b) $m_{H^{\pm}}=400$ GeV.  The leading $m_t^4$ radiative
correction of Eqs.~(\protect\ref{deltamassmatrix}) and
(\protect\ref{epsilon}) is included with $m_{\tilde{t}}= 1$ TeV in
calculating the remaining Higgs masses and mixings. The running quark
mass $m_b = 3.2$ GeV has been used, and all SUSY decay modes are
assumed suppressed.
\label{fig:br1}}
\end{figure}

\begin{figure}
\vspace*{.5in}\centerline{
\psfig{file=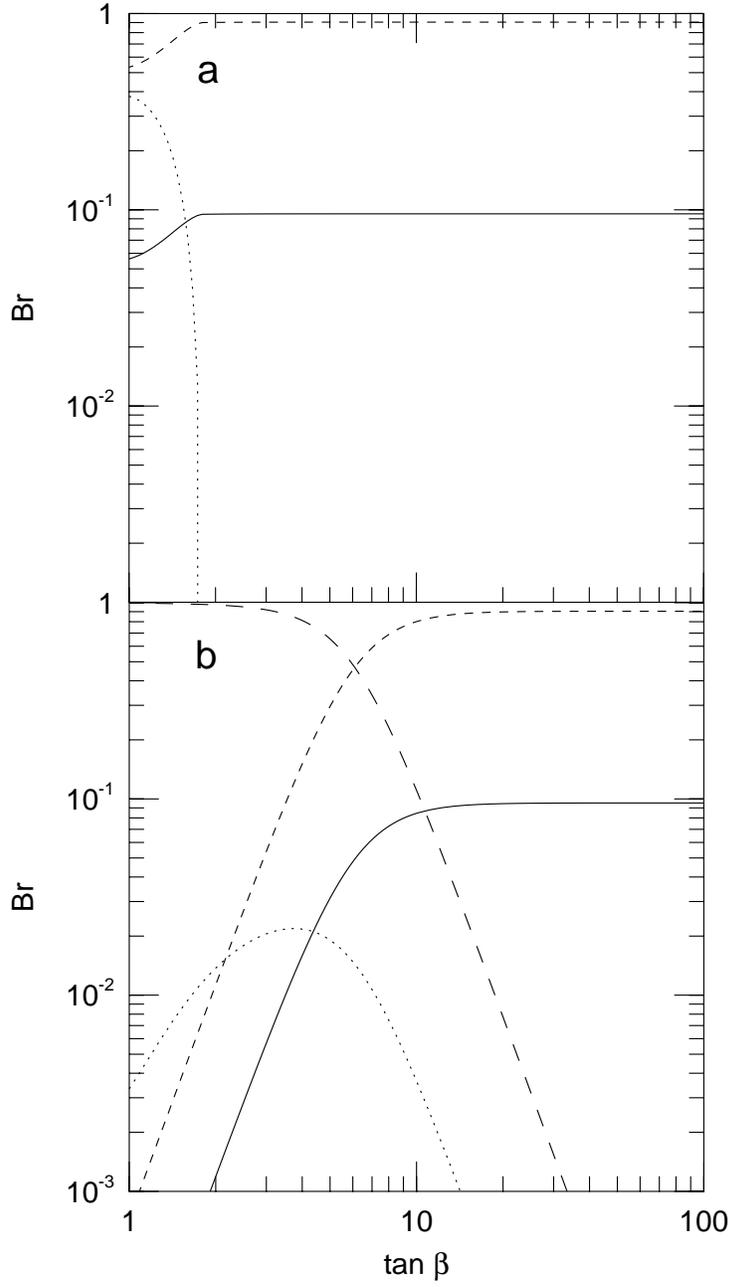,width=0.61\textwidth}
\vspace*{.5in}}
\caption{ Same as in Fig.~\protect\ref{fig:br1}, but for CP--odd
scalar $A$ decays to $\tau\bar{\tau}$ (solid), $b\bar{b}$ (short
dash), $t\bar{t}$ (long dash), and $Zh$ (dotted).
\label{fig:br2}}
\end{figure}

\begin{figure}
\vspace*{.5in}\centerline{
\psfig{file=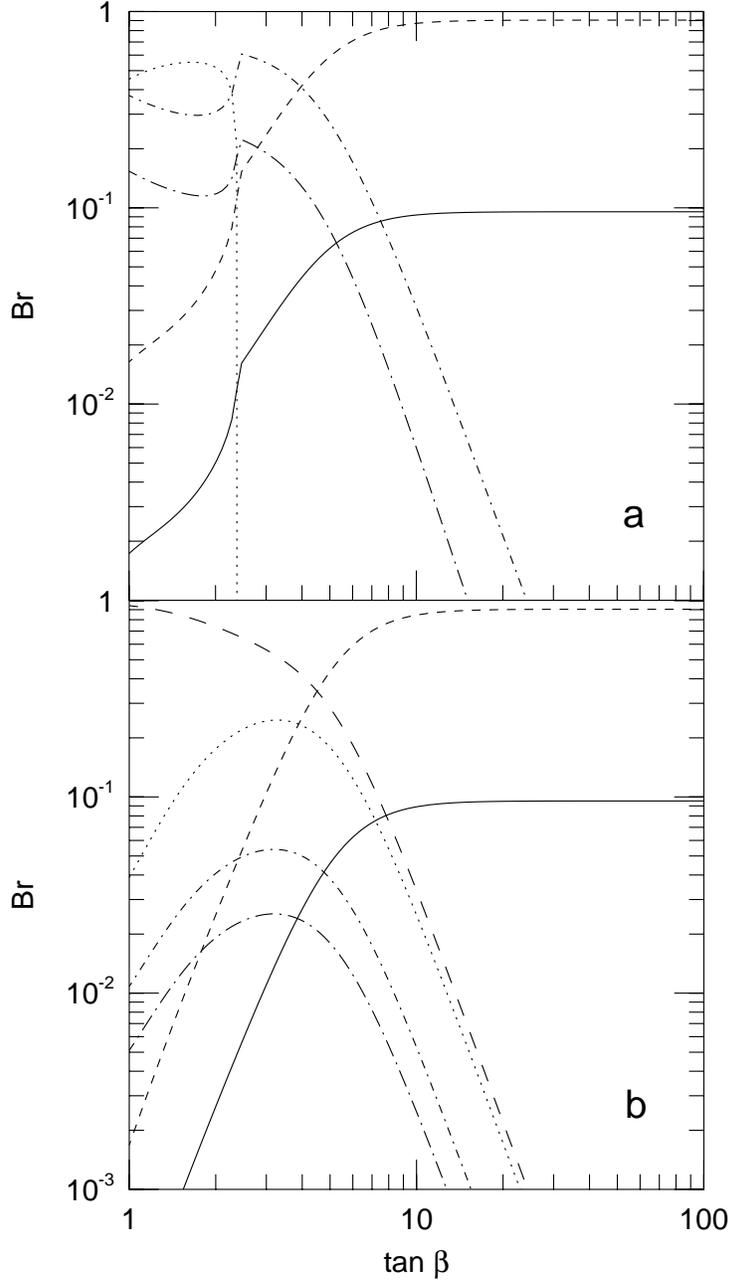,width=0.61\textwidth}
\vspace*{.5in}}
\caption{ Same as in Fig.~\protect\ref{fig:br1}, but for neutral Higgs
scalar $H$ decays to $\tau\bar{\tau}$ (solid), $b\bar{b}$ (short
dash), $t\bar{t}$ (long dash), $hh$ (dotted), $W^+W^-$ (dot -- short
dash), and $ZZ$ (dot -- long dash).
\label{fig:br3}}
\end{figure}

Although the two-body production cross sections are nearly independent
of $\tan\beta$, the heavy Higgs branching ratios are very sensitive to
$\tan\beta$.  The decay width formulas are given in the Appendix and
the branching ratios for $H^{\pm}$, $A$, and $H$ are plotted as
functions of $\tan\beta$ in Figs.~\ref{fig:br1}, \ref{fig:br2}, and
\ref{fig:br3}, respectively.  (Insignificant modes with branching
ratios never greater than $10^{-2}$ are omitted.)  In these plots, we
fix $m_{H^{\pm}}$.  The other Higgs masses and mixings are then
determined as functions of $\tan\beta$, including the leading $m_t^4$
radiative correction of Eqs.~(\ref{deltamassmatrix}) and
(\ref{epsilon}) with $m_{\tilde{t}}= 1$ TeV.  Note also that we use
$m_t = 175$ GeV and the running mass $m_b = 3.2$ GeV for the dynamical
coupling.  We have assumed that all SUSY decay modes are suppressed,
either kinematically or through mixing angles.  SUSY decay modes and
their effect on our analysis will be discussed in detail in
Sec.~\ref{sec:SUSYdecays}.

Several features of the branching ratios are important.  Throughout
this study, we assume $m_{H^{\pm}} > m_t+m_b$, so the decay $H^{\pm}
\to tb$ is always open.  Given that the decay widths are governed by
Yukawa couplings, one might expect that this decay mode would be
dominant for all values of $\tan\beta$. In fact, however, if charged
Higgses can be pair produced at $\sqrt{s} = 500$ GeV, the phase space
suppression for this decay is large, and the branching ratio for $H^-
\to \tau \bar{\nu}$ can be substantial, as may be seen in
Fig.~\ref{fig:br1}a.  Charged Higgs events with mixed decays, $H^+H^-
\to t\bar{b}\tau \bar{\nu}, \bar{t} b \bar{\tau} \nu$, will be very
useful for determining $\tan\beta$.  We will see that when the
branching ratio $B(H^- \to \tau \bar{\nu})$ depends strongly on
$\tan\beta$, roughly for $3 \alt \tan\beta \alt 10$, we will be able
to determine $\tan\beta$ precisely from this channel.  For
$m_{H^{\pm}} = 400$ GeV, as can be seen in Fig.~\ref{fig:br1}b, the
$tb$ phase space suppression is negligible, and the ratio
$\Gamma(\tau\bar{\nu})/\Gamma(\bar{t} b)$ approaches $m_{\tau}^2 / 3
m_b^2 \approx 0.1$ for large $\tan\beta$.

For the neutral Higgs bosons, two features are particularly
noteworthy.  For $m_{H^{\pm}} = 200$ GeV, $t\bar{t}$ decays are
closed, and we see in Figs.~\ref{fig:br2}a and \ref{fig:br3}a that
decays to Higgs and gauge bosons are substantial for low $\tan\beta$.
Such modes will result in 4$b$, 6$b$, and even 8$b$ events from $AH$
production, leading to distinct signals.  These branching ratios
decrease rapidly as $\tan\beta$ increases through moderate values, so
again we expect strong determinations of $\tan\beta$ in this region.
For $m_{H^{\pm}} = 400$ GeV, the $t\bar{t}$ mode is open and
completely dominates for low $\tan\beta$, as we see in
Figs.~\ref{fig:br2}b and \ref{fig:br3}b.  However, as $\tan\beta$
increases, this mode is suppressed by $\tan^2\beta$, while the
$b\bar{b}$ mode is enhanced by $\tan^2\beta$.  Ignoring phase space
suppressions, the branching ratios cross roughly at $\tan\beta \approx
(m_t/m_b)^{1/2} \approx 7$, so again we expect a strong determination
of $\tan\beta$ when its underlying value is in the middle range.

\section{Experimental Simulation}
\label{sec:Experiment}

\subsection{Signals and Cuts}
\label{subsec:Channels}

As seen in Sec.~\ref{sec:Tanb}, Higgs production leads to events with
many $b$ quarks.  We will exploit this feature in conjunction with the
excellent $b$-tagging efficiency, $\epsilon_b \approx 60\%$, that is
expected to be available at future $e^+e^-$ colliders\cite{Jackson}.
As many decay channels may be open, Higgs production contributes to
many types of events.  In this section, we first list the eight signal
channels that we will consider. We then return to them in detail, and
explain our motivations for choosing them.

We consider the following channels:
\begin{itemize}

\item[{\bf 1}] $2b+l+q\text{'s} + \text{ cuts 1a -- 1e below}$ 
(``$H^+H^-$'' channel).

\item[{\bf 2}] $2b+l+q\text{'s} + \text{ cuts 2a -- 2e below}$ 
(``$tbH^{\pm}$'' channel).

\item[{\bf 3}] $3b+1l\ (+q\text{'s})$.

\item[{\bf 4}] $3b+0,2,3,\ldots\ l\ (+q\text{'s})$.

\item[{\bf 5}] $4b$.

\item[{\bf 6}] $4b+1l\ (+q\text{'s})$.

\item[{\bf 7}] $4b+ 0,2,3,\ldots\ l\  (+q\text{'s})$ (but not $4b$). 

\item[{\bf 8}] $5b\ (+\ l+q\text{'s})$.

\end{itemize} 
In this list, ``$b$'' and ``$q$'' denote hadronic jets with and
without a $b$ tag, respectively, ``$l$'' denotes an isolated,
energetic $e$, $\mu$, or $\tau$, and particles enclosed in parentheses
are optional. In our analysis, we assume that hadronically-decaying
$\tau$ leptons may be identified as leptons, ignoring the slight
degradation in statistics from multi-prong $\tau$ decays.  Thus, for
example, channel 6 contains all events with 4 tagged $b$'s and exactly
one $e$, $\mu$ or $\tau$, with or without untagged jets. We defer the
details of the implementation of these requirements in our simulations
to Sec.~\ref{subsec:Simulation}.

\begin{figure}
\vspace*{.5in}\centerline{
\psfig{file=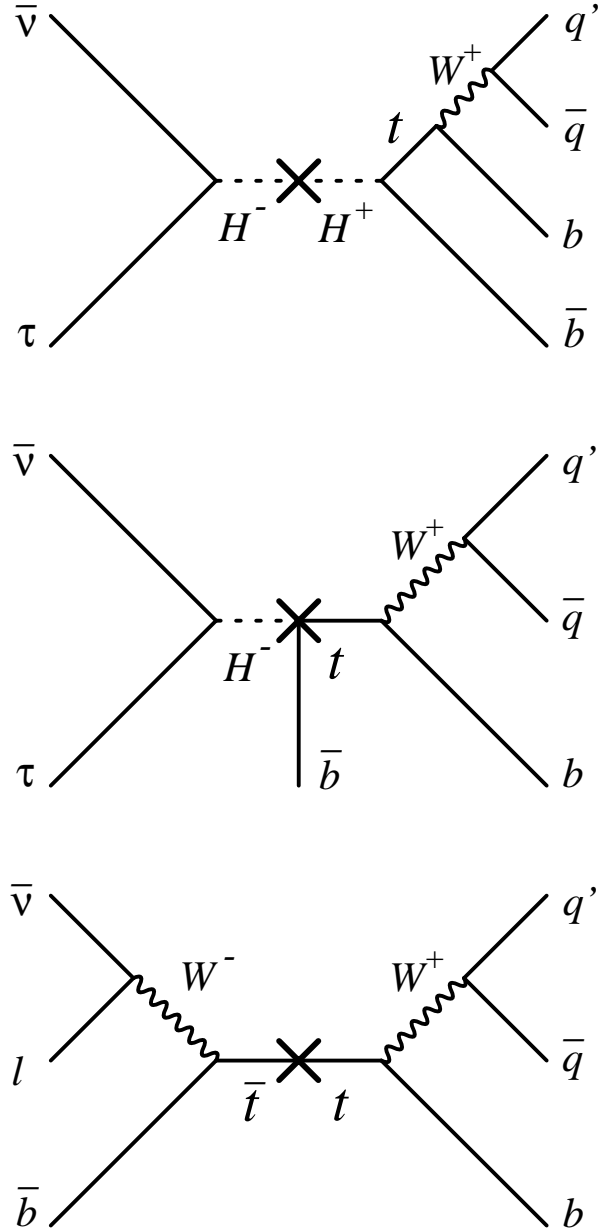,width=0.55\textwidth}
\vspace*{.5in}}
\caption{ Schematic pictures of signal events $H^+H^- \to
t\bar{b}\tau\bar{\nu}$ and $t\bar{b}H^- \to
t\bar{b}\tau\bar{\nu}$, and background $t\bar{t} \to b \bar{q} q'
\bar{b} l \bar{\nu}$.  The crosses mark the interaction point.
\label{fig:eventpicture}}
\end{figure}

We now discuss each channel in detail.  Let us begin by discussing
channels 1 and 2, which are intended to isolate $e^+e^-\rightarrow
H^+H^-\rightarrow tb\tau\nu$ and $e^+e^-\rightarrow tbH^\pm\rightarrow
tb\tau\nu$, respectively. The cross sections for these events are
strongly dependent on $\tan\beta$: as seen in Fig.~\ref{fig:br1}, the
$H^{\pm}$ branching fractions vary rapidly for moderate $\tan\beta$,
and we will see below that the cross section for $tbH^{\pm}$ grows
rapidly for large $\tan\beta$. These channels therefore allow us to
bound $\tan\beta$ if we can reduce backgrounds to low levels.  The
event shapes of these two channels and the largest background
$e^+e^-\rightarrow t\bar{t}\rightarrow tbl\nu$ are given schematically
in Fig.~\ref{fig:eventpicture}.  We see that all three processes have
exactly the same final state particle content: $2b+q\text{'s}$, with
an isolated single lepton.  The cross section for $t\bar{t}$
production is about one order of magnitude larger than that of
$H^+H^-$ production, and hence it is crucial that this background (as
well as others) be reduced by additional cuts.

To reduce backgrounds in both channels, we first attempt to
reconstruct the $W$ boson and top quark. Each candidate event has an
untagged hadronic system from $W^\pm$ decay, which we denote ``had'',
and two $b$ jets, one of which comes from $t$ decay. Thus, we first
rescale $p_{\text{had}}$ to obtain the correct value of the $W$ boson
mass:

\begin{equation}
(r_W p_{\text{had}})^2 = m_W^2 \ ,
\end{equation}
where $p_{\text{had}}$ is the four-momentum of the untagged hadronic
system measured in the detector, and $r_W$ is the rescaling factor
defined by this equation. In the signal event, $r_W$ is close to 1.
Then, we try to reconstruct the top quark mass from the four-momentum
$r_W p_{\text{had}}$ and one of the $b$ momenta.  However, if the $b$
from $t$ decay decays semi-leptonically, the neutrino carries away
some fraction of the momentum.  We therefore also define, for each $b$
jet, a rescaling factor $r_b$ given by

\begin{equation}
(r_b p_{b} + r_W p_{\text{had}})^2 = m_t^2 \ ,
\end{equation}
where $p_b$ is the four-momentum of the $b$ jet. For the $b$ coming
from the $t$ decay, $r_b$ is almost 1 in an event without
semi-leptonic decay. If the $b$ decays semi-leptonically, the neutrino
typically carries away about 25\% of the total energy in the $b$ rest
frame\cite{semi-leptonic}, and $r_b$ may become larger than 1.
However, even in this case, $r_b$ is less than 1.7 for about 90\% of
the events. We therefore require that, for at least one of the $b$
jets, $0.9 \le r_b \le 1.7$.  We then identify the $b$ jet most likely
to have come from $t$ decay as $b_1$ and the other as $b_2$.  This is
done as follows: if only one $r_b$ is in the range $0.9 \le r_b \le
1.7$, we define the $b$ jet corresponding to this $r_b$ as $b_1$, and
the other as $b_2$.  If both $r_b$'s are in this interval, we identify
the $b$ with $r_b$ closer to 1 as $b_1$, and the other as $b_2$. The
untagged hadronic system and $b_1$ then form the candidate top quark
system.

Given these definitions of $r_W$, $r_b$, $b_1$, and $b_2$, we then
impose several kinematic cuts. For channel 1, this is quite simple. In
the $H^+H^-$ pair production event, the untagged hadronic system and
the 2 $b$'s are all decay products of one $H^{\pm}$, and hence their
total energy is (in principle) equal to the beam energy
$\frac{1}{2}\sqrt{s}$. On the other hand, the total hadronic system of
the $t\bar{t}$ background has more energy, since the decay products of
one top quark alone already have energy equal to the beam energy.
Thus, we require that the energy of the total hadronic system be
approximately the beam energy: $\frac{1}{2}\sqrt{s}-\Delta
E_{H^{\pm}}^- \le r_W E_{\text{had}} + r_{b_1} E_{b_1} + E_{b_2} \le
\frac{1}{2}\sqrt{s}+\Delta E_{H^{\pm}}^+$.  (The numerical values of
the cut parameters $\Delta E_{H^{\pm}}^{\pm}$ and those that follow
depend on $\sqrt{s}$; they are given in Table~\ref{table:cut_params}.)
Furthermore, a cut on the energy of the candidate top quark system also
effectively reduces the $t\bar{t}$ background; we require
$r_WE_{\text{had}} + r_{b_1} E_{b_1} \le \frac{1}{2}\sqrt{s}-\Delta
E_t^{\rm 1b}$. We also impose a cut on the invariant mass of the bottom
quark pair $m_{b_1b_2}$ to eliminate backgrounds in which the $b$ jets
arise from $Z$ boson decay ({\em e.g.}, $e^+e^-\rightarrow W^+W^-Z$):
$m_{b_1b_2}\le m_Z-\Delta m_Z^-$ or $m_{b_1b_2}\ge m_Z+\Delta m_Z^+$.
Finally, we require that the invariant mass of the untagged jets satisfy
$|m_{\text{had}}-m_W| \le\Delta m_W$, and that the single lepton have
energy $E_l>5$ GeV and be isolated with no hadronic activity within a
cone of half-angle $20^{\circ}$.  In summary, for channel 1, we adopt
the following cuts: 
\begin{itemize}

\item[{\bf 1a}] $\frac{1}{2}\sqrt{s}-\Delta E_{H^\pm}^{-}
\le r_W E_{\text{had}} + r_{b_1} E_{b_1} + E_{b_2} \le
\frac{1}{2}\sqrt{s}+\Delta E_{H^\pm}^{+}$, with 
$0.9 \le r_{b_1} \le 1.7$.

\item[{\bf 1b}] $r_WE_{\text{had}} + r_{b_1} E_{b_1} \le
\frac{1}{2}\sqrt{s}-\Delta E_t^{\rm 1b}$.

\item[{\bf 1c}] $m_{b_1b_2}\le m_Z-\Delta m_Z^{-}$ or $m_{b_1b_2}\ge
m_Z+\Delta m_Z^{+}$.

\item[{\bf 1d}] $|m_{\text{had}}-m_W|\le\Delta m_W$.

\item[{\bf 1e}] The single lepton must be energetic, $E_l>5$ GeV, and
isolated, with no hadronic activity within a cone of half-angle
$20^{\circ}$.

\end{itemize}

The situation for channel 2 (the ``$tbH^{\pm}$'' channel) is more
complicated, since on-shell production of $H^+H^-$ pairs is also a
background. For this reason, we replace cuts 1a and 1b above.  In
order to make sure that we have only one ``on-shell'' $H^{\pm}$, we
impose a cut on the energy of the total hadronic system:
$r_WE_{\text{had}} + r_{b_1} E_{b_1} + E_{b_2}\ge
\frac{1}{2}\sqrt{s}+\Delta E_{H^\pm}^{+}$. This cut does not
effectively eliminate the $t\bar{t}$ background, in contrast to cut
1a, and we must thus rely on a cut on the energy of the candidate top
quark system to remove the $t\bar{t}$ background. In order to reduce
the $t\bar{t}$ background even if we misidentify the $b$ jet, we
modify cut 1b and instead require that $r_WE_{\text{had}} + r_{b_1}
E_{b_1}\le\frac{1}{2}\sqrt{s}-\Delta E_t^{\rm 2b}$ and
$r_WE_{\text{had}} + r_{b_2} E_{b_2} \le \frac{1}{2}\sqrt{s}-\Delta
E_t^{\rm 2b}$.  In addition, we again use cuts 1c -- 1e.  The
following is then the complete set of cuts applied to channel 2:
\begin{itemize}

\item[{\bf 2a}] $r_WE_{\text{had}} + r_{b_1} E_{b_1} + E_{b_2} \ge
\frac{1}{2}\sqrt{s}+\Delta E_{H^\pm}^{+}$, with
$0.9 \le r_{b_1} \le 1.7$.

\item[{\bf 2b}] $r_WE_{\text{had}} + r_{b_1}
E_{b_1}\le\frac{1}{2}\sqrt{s}- \Delta E_t^{\rm 2b}$ and
$r_WE_{\text{had}} + r_{b_2} E_{b_2} \le \frac{1}{2}\sqrt{s}-\Delta
E_t^{\rm 2b}$.

\item[{\bf 2c}] $m_{b_1b_2}\le m_Z-\Delta m_Z^{-}$ or $m_{b_1b_2}\ge
m_Z+\Delta m_Z^{+}$.

\item[{\bf 2d}] $|m_{\text{had}}-m_W|\le\Delta m_W$.

\item[{\bf 2e}] The single lepton must be energetic, $E_l>5$ GeV, and
isolated, with no hadronic activity within a cone of half-angle
$20^{\circ}$.

\end{itemize} 

The choice of channels 3 -- 8 is motivated by a number of
considerations.  To exploit the $b$-rich events in Higgs signals, we
require many $b$ tags.  For channels 3 -- 8, the requirement of three
to five $b$ tags effectively removes most standard model backgrounds.
In fact, $AH$ production may result in events with as many as eight
$b$ quarks, and so channels requiring more than five $b$ tags may also
be considered.  However, once branching ratios and $b$-tagging
efficiencies are included, such channels suffer from poor statistics
and do not improve our results.

In channels 3 -- 7, events with exactly 1 lepton are distinguished
from the others. Charged Higgs interactions are not plagued by large
1--loop corrections, and therefore provide signals that are not
subject to large systematic errors.  On the other hand, $AH$ events
may be subject to such uncertainties, in particular in the interaction
vertex $Hhh$.  (See Sec.~\ref{sec:Systematics}.)  To avoid
contaminating the charged Higgs signal with systematic uncertainties,
we would like to separate the $H^+ H^-$ and $AH$ events.  This may be
achieved for some parameters by separating $1l$ events, as charged
Higgs pair production may produce $1l$ events through $H^+H^- \to
t\bar{b}\bar{t} b \to b\bar{b}b\bar{b}W^+W^-$, where one $W$ decays
leptonically and the other hadronically.  $AH$ events generally do
not, unless $A, H\to t \bar{t}$ decays are open.

\subsection{Signal Simulation}
\label{subsec:Simulation}

We must now determine the size of the signal in each channel after all
branching ratios, cuts, and tagging efficiencies are included.  The
cross sections in channels 3 -- 8 are completely determined by
branching ratios and the $b$-tagging efficiency.  In channels 1 and 2,
where kinematic cuts apply, we must simulate each of the signal
events.  Events were generated with a parton level Monte Carlo event
generator, using the helicity amplitude package HELAS\cite{HELAS} and
phase space sampler BASES\cite{BASES}. For both cases, the spin
correlations present in the decays were not included.  Semi-leptonic
$b$ decays were simulated with branching fraction 24\% and energy
distribution given in Ref.\cite{semi-leptonic}.  As stated previously,
we use the running quark mass values $m_t = 175$ GeV and $m_b = 3.2$
GeV in branching fraction calculations.
 
Hadronization and detector effects were crudely simulated by smearing
the parton energies with detector resolutions projected to be
available at future $e^+e^-$ colliders\cite{JLCI}: $\sigma_E^{\text{
had}}/E = 40\%/\sqrt{E} \oplus 2\%$, $\sigma_E^{\text{ e.m.}}/E =
15\%/\sqrt{E} \oplus 1\%$, and $\sigma_{p_{\perp}}/p_{\perp} = 5\times
10^{-4} p_{\perp} \oplus 0.1\%$ for muons, with $E$ and $p_{\perp}$ in
GeV.  The efficiency of cut 2a is quite sensitive to the hadronic
calorimeter resolution, but we assume that the characteristics of this
calorimeter are well-understood.  For the purpose of imposing the
cuts, an isolated lepton was defined to be one with no hadronic parton
within a cone of half-angle $20^{\circ}$.  In addition, we assume that
$\tau$ leptons may be identified as leptons.  Finally, initial state
radiation was not included. For $\sqrt{s} = 1$ TeV, the effects of
initial state radiation may be substantial, and the cuts we have
proposed may require modification.

The number of signal and background events in each channel is heavily
influenced by the $b$-tagging efficiency and purity.  Recently, there
have been great improvements in this area.  In this study, we assume
that the probability of tagging a $b$ ($c$) quark as a $b$ quark is
$\epsilon_b=60\%$ ($\epsilon_c=2.6\%$)\cite{Jackson}.  (The
possibility of tagging light quarks as $b$ quarks is negligible.)
Uncertainties in these parameters will contribute to our systematic
errors (see Sec.~\ref{sec:Systematics}).  In addition, we present
results for other $b$-tagging efficiencies in Sec.~\ref{sec:Results}.

\begin{figure}
\vspace*{.5in}\centerline{
\psfig{file=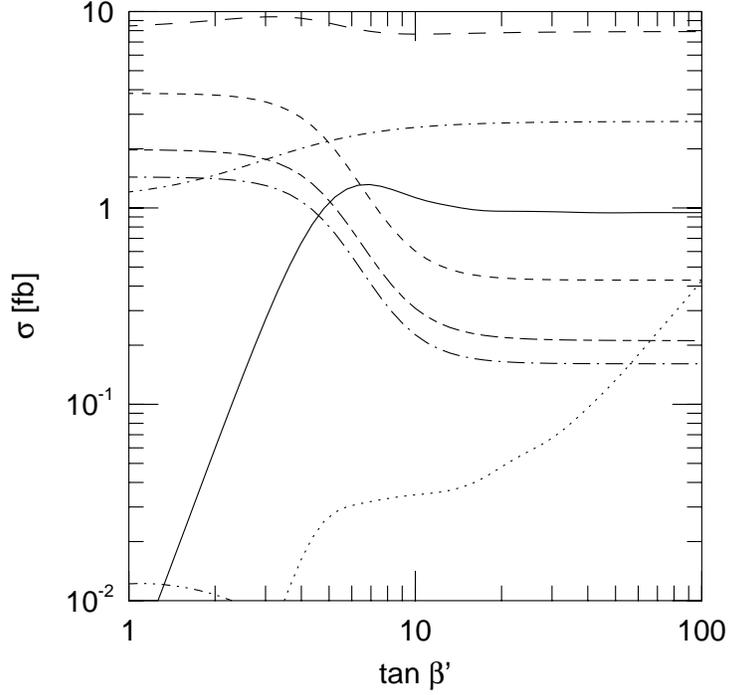,width=0.61\textwidth}
\vspace*{.5in}}
\caption{ Signal cross sections for $m_{H^{\pm}}=200$ GeV,
$\tan\beta=5$, and $\protect\sqrt{s}= 500$ GeV, in each channel: (1)
$2b+l+q\text{'s} + \text{ cuts 1a--1e}$ (``$H^+H^-$'' channel)
(solid), (2) $2b+l+q\text{'s} + \text{ cuts 2a--2e}$ (``$tbH^{\pm}$''
channel) (dotted), (3) $3b+1l\ (+q\text{'s})$ (short dash), (4)
$3b+0,2,3,\ldots\ l\ (+q\text{'s})$ (long dash), (5) $4b$ (dot--short
dash), 6) $4b+1l\ (+q\text{'s})$ (dot--long dash), (7) $4b+
0,2,3,\ldots\ l\ (+q\text{'s})$ (but not $4b$) (short dash--long
dash), and (8) $5b\ (+\ l+q\text{'s})$ (dash--dot--dot).  All
branching fractions, tagging efficiencies, and cuts have been
included.
\label{fig:channels1}}
\end{figure}

\begin{figure}
\vspace*{.5in}\centerline{
\psfig{file=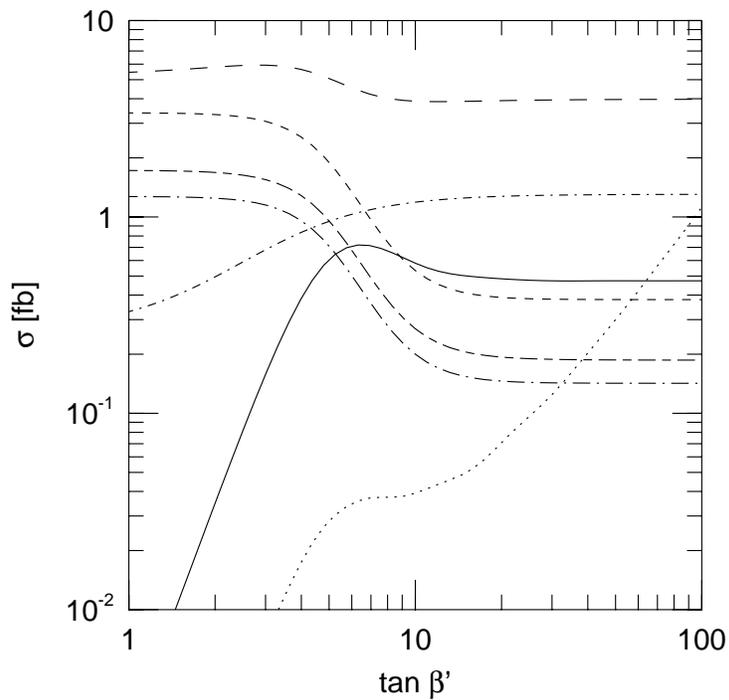,width=0.61\textwidth}
\vspace*{.5in}}
\caption{ Signal cross sections as in
Fig.~\protect\ref{fig:channels1}, but for $m_{H^{\pm}}=200$ GeV and
$\protect\sqrt{s}= 1$ TeV. The cross section for channel 8 is less
than $10^{-2}\text{ fb}$, and therefore does not appear.
\label{fig:channels2}}
\end{figure}

\begin{figure}
\vspace*{.5in}
\centerline{
\psfig{file=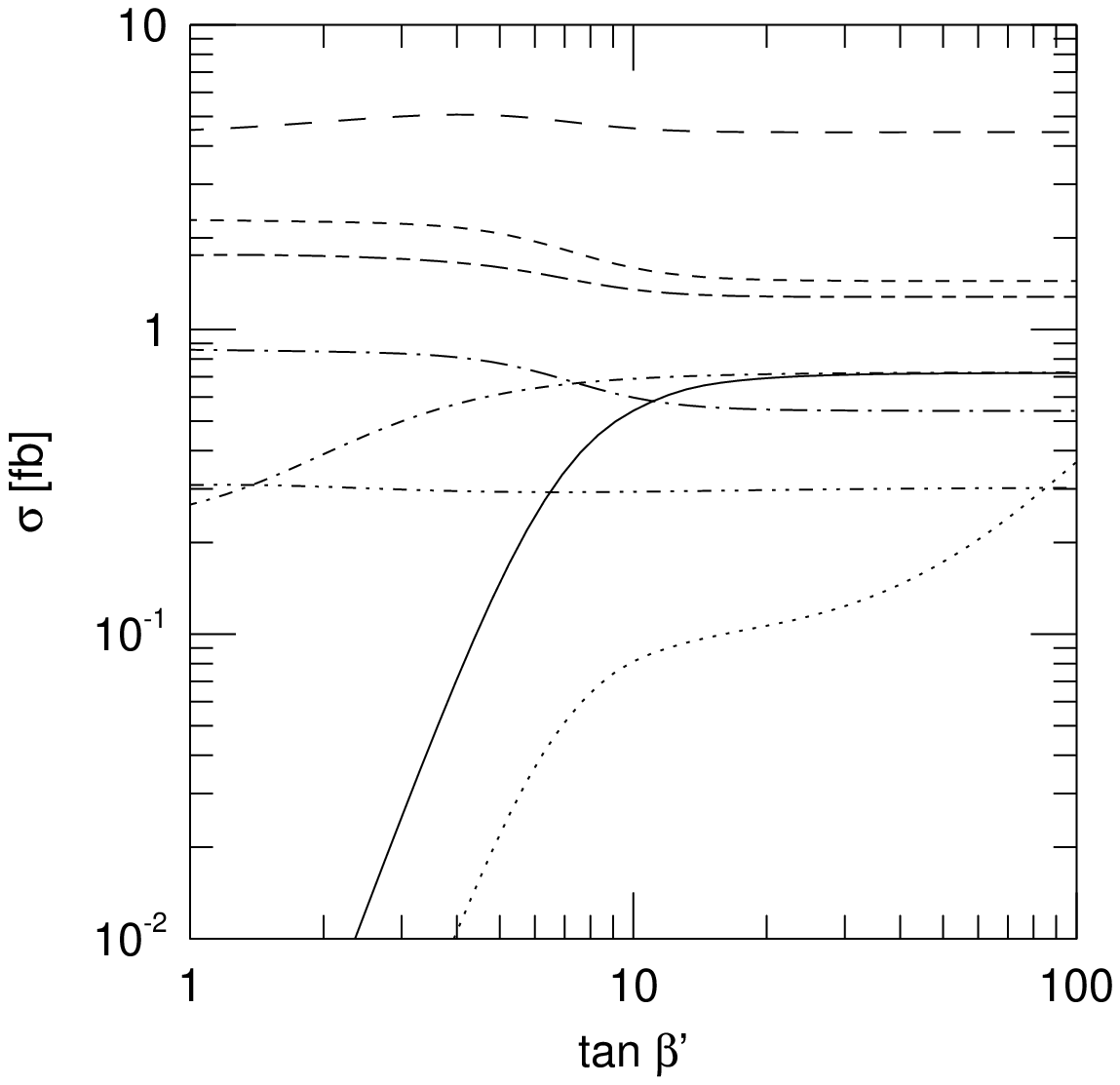,width=0.61\textwidth}
\vspace*{.5in}}
\caption{
Signal cross sections as in Fig.~\protect\ref{fig:channels1}, but for
$m_{H^{\pm}}=300$ GeV and $\protect\sqrt{s}= 1$ TeV.
\label{fig:channels3}}
\end{figure}

\begin{figure}
\vspace*{.5in}
\centerline{
\psfig{file=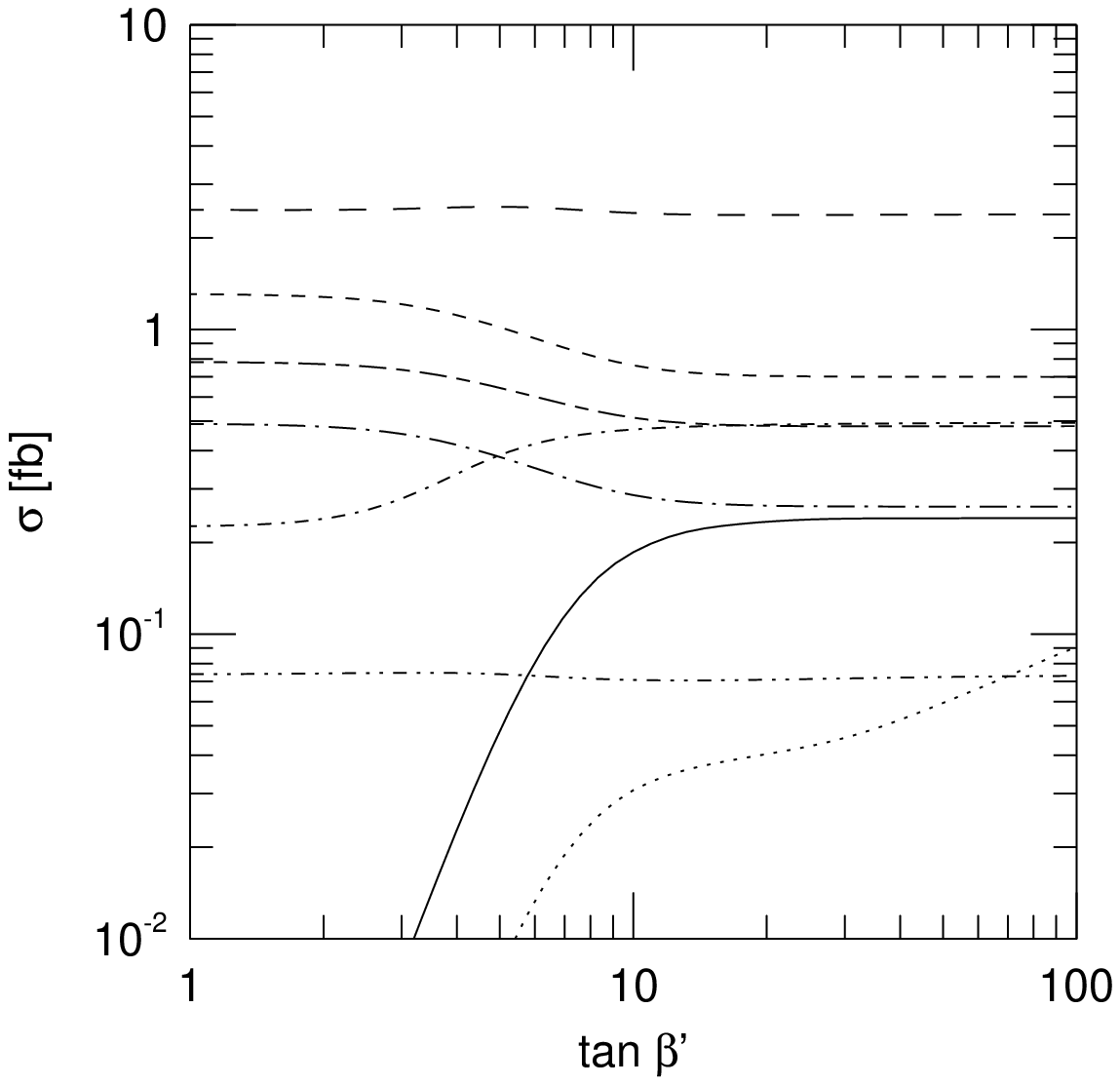,width=0.61\textwidth}
\vspace*{.5in}}
\caption{
Signal cross sections as in Fig.~\protect\ref{fig:channels1}, but for 
$m_{H^{\pm}}=400$ GeV and $\protect\sqrt{s}= 1$ TeV.
\label{fig:channels4}}
\end{figure}

With these assumptions, we can now present the expected signal cross
sections after cuts in each channel.  In Figs.~\ref{fig:channels1} --
\ref{fig:channels4}, we display these cross sections as functions of a
postulated $\tan\beta'$ for fixed underlying values $\tan\beta=5$ and
$(\sqrt{s}, m_{H^{\pm}}) =$ (500 GeV, 200 GeV), (1 TeV, 200 GeV), (1
TeV, 300 GeV) and (1 TeV, 400 GeV).  These figures are generated as
follows: we first assume that the underlying values of $\tan\beta$ and
$m_{H^{\pm}}$ realized in nature are as given above. We then calculate
the Higgs masses and compositions, including the radiative corrections
of Eqs.~(\ref{deltamassmatrix}) and (\ref{epsilon}) with fixed
$m_{\tilde{t}}=1$ TeV, as discussed in Sec.~\ref{subsec:Definitions}.
In particular, this determines the values of the physical Higgs
masses, $\cos^2 (\beta-\alpha )$, and $B(H\to hh)$ that would
presumably be measured.  With these quantities then held fixed to
their measured values, we then consider a hypothetical $\tan\beta'$,
and determine the contributions of the $H^+H^-$, $AH$, $Zh$, and
$tbH^{\pm}$ signals to the various channels after cuts as a function
of $\tan\beta'$.  The contributions to channels 3 -- 8 are determined
simply by branching ratios and tagging efficiencies.  $Zh$ events do
not contribute to channels 1 and 2, as they never have exactly one
lepton.  The contributions of the other 3 processes to channels 1 and
2 are determined by Monte Carlo simulation. The signal efficiency of
the cuts for channel 1 are 48\% -- 34\% (31\% -- 19\%, 55\% -- 62\%,
and 60\% -- 64\%) for $\sqrt{s}=500$ GeV with $m_{H^{\pm}}=200$ GeV
($\sqrt{s}=1$ TeV with $m_{H^{\pm}}=200$, 300, and 400 GeV), as we
vary $\tan\beta'$ from 1 to 100. For channel 2, the efficiency is 33\%
(49\%, 79\%, and 83\%) for $\sqrt{s}=500$ GeV with $m_{H^{\pm}}=200$
GeV ($\sqrt{s}=1$ TeV with $m_{H^{\pm}}=200$, 300, and 400 GeV) and
$\tan\beta' =60$.

We see in Figs.~\ref{fig:channels1} -- \ref{fig:channels4} the
expected behavior, given the branching ratios shown in
Figs.~\ref{fig:br1}, \ref{fig:br2}, and \ref{fig:br3}.  For
$m_{H^{\pm}} = 200$ GeV, we see in Figs.~\ref{fig:channels1} and
\ref{fig:channels2} that channels 3 -- 8 are large for low
$\tan\beta$, and drop rapidly for increasing values of $\tan\beta$.
The cross sections in these channels are enhanced when the $A\to Zh$
and $H\to hh$ branching fractions are large, as is the case for low
$\tan\beta$.  In addition, we see that the cross section in channel 1
grows rapidly from low $\tan\beta$ to high $\tan\beta$, as the
branching ratio for $H^- \to \tau \bar{\nu}$ grows.  Finally, the
cross section for $tbH^{\pm}$ is virtually non-existent for low
$\tan\beta$, but grows rapidly for $\tan\beta \agt 20$.  For
$\tan\beta \approx 60$, we see that the cross section is large enough
to produce tens of events per year, allowing a promising determination
for high $\tan\beta$ if the backgrounds are small.

Each of the dependencies on $\tan\beta$ mentioned above is weakened
for larger $m_{H^{\pm}}$, as may be seen first for $m_{H^{\pm}}=300$
GeV in Fig.~\ref{fig:channels3}.  For $m_{H^{\pm}} = 400$ GeV, the
$t\bar{t}$ decay mode is now open.  The branching fractions of $A\to
Zh$ and $H\to hh$ are thus not very large even for low $\tan\beta$,
and so the dependence on $\tan\beta$ of channels 3 -- 8, though still
present, is diluted.  The cross section for channel 1 also does not
rise as much as $\tan\beta$ increases, as the cross section $\sigma (
H^+ H^- \to tb\tau\nu)$ is now no longer enhanced by the phase space
suppression of $H^{\pm} \to tb$, as it was in the $m_{H^{\pm}} = 200$
GeV case. Furthermore, because of the smaller branching ratio for
$H^{\pm} \to \tau\nu$, as well as the phase space suppression for the
production process $e^+e^-\rightarrow tbH^{\pm}$, the cross section
for channel 2 at high $\tan\beta$ is not as large as in the other
cases.

\subsection{Backgrounds}

In our analysis, we have included the following standard model
backgrounds (cross sections in fb before cuts at $\sqrt{s} = 500$ GeV
and 1 TeV are given in parentheses): $t\bar{t}$ (540, 180), $W^+W^-$
(7000, 2700), $ZZ$ (400, 150), $t\bar{t}Z$ (1.2, 4.7), $W^+W^-Z$ (40,
60), $ZZZ$ (1.0, 0.85), $\nu\bar{\nu} t\bar{t}$ (0.01, 0.55), $e^+e^-
t\bar{t}$ (0.35, 6.0), $\nu\bar{\nu} ZZ$ (0.6, 6.5), $e^+e^- ZZ$ (1.0,
2.5), $e^+e^- W^+ W^-$ (250, 1100), $\nu\bar{\nu} W^+W^-$ (2.0, 16),
$e\nu WZ$ (8.0, 70), and $t\bar{t}h$ (2.0, 3.5).  The cross sections
for all but the last of these processes have been calculated in
Ref.~\cite{HMthesis}, and cross sections for the last process may be
found in Ref.~\cite{DKZ}.  The cross section for $t\bar{t}h$, in fact,
depends on $\tan\beta$.  However, as this cross section is small, the
influence of this dependence is rather weak, and we treat it as
constant.

The contributions of these backgrounds to channels 3 -- 8 are
completely determined by their cross sections, branching fractions,
and $\epsilon_b$ and $\epsilon_c$. The estimated background cross
sections for these channels, as well as those for channels 1 and 2,
are given in Table~\ref{table:bg}.  For channels 1 and 2, the
kinematic requirements of Sec.~\ref{subsec:Channels} must be imposed.
By far the largest background to these channels before cuts is
$t\bar{t}$ production.  To obtain an accurate estimate of the
contribution of this background after cuts, we have simulated
$t\bar{t}$ events using the Monte Carlo program described above,
neglecting production-decay spin correlations.  For $\sqrt{s} = 500$
GeV, the signal and background efficiencies as each additional cut is
applied are given in Tables~\ref{table:1} and \ref{table:2} for
channels 1 and 2. The cuts are seen to be excellent for removing the
$t\bar{t}$ background, reducing this background to 0.046 fb (channel
1) and 0.012 fb (channel 2). For $\sqrt{s} = 1$ TeV, the efficiencies
for the $t\bar{t}$ background are also very small ($1.7\times 10^{-3}$
for channel 1, and $1.4\times 10^{-3}$ for channel 2), and the
background is again well suppressed (0.045 fb for channel 1, and 0.036
fb for channel 2).

Given the effectiveness of the cuts for the $t\bar{t}$ background, we
next consider other backgrounds.  We begin with the backgrounds for
$\sqrt{s} = 500$ GeV.  The processes $WWZ \to l\nu q' \bar{q} bb$ and
$e\nu WZ \to e\nu q' \bar{q} bb$ (when the $e$ does {\em not} go down
the beampipe) are possible backgrounds.  They are, however,
effectively removed by the kinematic cuts, especially 1c and 2c, which
are designed to eliminate events in which both $b$ quarks originate
from the decay of a $Z$ boson.  After the cuts, these backgrounds are
negligible.  The background $t\bar{t}g \to tbl\nu g$ generically fails
cuts 1d and 2d, as the untagged hadronic system consists of the gluon
jet and also the hadronic decay products of the $W$, and therefore
typically has invariant mass greater than $m_W$.  For this background
to pass the cuts, the gluon jet must mistakenly be included in one of
the tagged $b$ jets, which greatly reduces the background.
Furthermore, even if cuts 1d and 2d fail to eliminate this background,
it may be removed by considering the invariant mass of the combined
lepton and missing momentum.  In the signal, this is $\sim
m_{H^{\pm}}$, whereas in the background it is $\sim m_W$.  We have not
included this background and this cut, but we expect the cut to
degrade our signal efficiency very slightly.

Events $t\bar{t}(h,Z) \to bl\nu bq' \bar{q} (h,Z)$ may contribute to
channels 1 and 2 when $(h,Z)$ decays hadronically and only two of the
six jets are $b$-tagged.  However, as in the case of $t\bar{t}g$, such
events will fail cuts 1d and 2d, as the hadronic system again consists
of the hadronic decay products of the $W$ and additional hadronic
jets.  Events $t\bar{t}Z$ where $Z\to \nu \bar{\nu}$ may pass the
cuts, however.  In addition, $e^+e^- t\bar{t}$ events may also pass
the cuts when both electrons are lost in the beampipe.  We have not
simulated these processes, but rather assume conservatively that the
kinematic cuts do not further reduce these backgrounds.  After
including branching ratios and tagging efficiencies, channels 1 and 2
combined receive contributions from $t\bar{t}(Z\to \nu \bar{\nu})$ and
$e^+e^- t\bar{t}$ of 0.038 and 0.056 fb, respectively.  We will
conservatively assume backgrounds of 0.06 fb in each channel.

For $\sqrt{s}=1$ TeV, $t\bar{t}(Z\to \nu \bar{\nu})$ may again pass
all cuts. After including branching ratios and tagging efficiencies,
this contributes 0.15 fb.  The process $e^+e^- t\bar{t}$ also
contributes, and its cross section is now greatly increased, but now
the $t\bar{t}$ system is often not energetic enough to pass cuts 1a
and 2a.  The $E_{t\bar{t}}$ distribution for $\sqrt{s}= 1.5$ TeV is
presented in Ref.~\cite{HMthesis}.  From this figure, we may
extrapolate to $\sqrt{s}=1$ TeV to estimate that roughly 40\% of the
$e^+e^- t\bar{t}$ events have $E_{t\bar{t}} > \sqrt{s}/2$.  (We have
checked that reasonable deviations from this value do not
significantly change the results presented below.) Including branching
ratios and tagging efficiencies and the 40\% efficiency, this
background is 0.38 fb after cuts. Finally, the $\nu\bar{\nu} t\bar{t}$
background is now non-negligible, and gives 0.088 fb, conservatively
including only branching ratios and tagging efficiencies.  In summary,
for $\sqrt{s}=1$ TeV, we estimate the total backgrounds to channels 1
and 2 combined to be 0.62 fb.  We assume backgrounds of 0.31 fb in
each channel.

$AH$ production may also contribute to channels 1 and 2 when the decay
$H\to W^+W^-$ is prominent.  Although a signal, this mode has a number
of systematic errors that would contaminate the $H^+H^-$ contribution
we are trying to isolate.  However, in our Monte Carlo simulation, we
find that for channel 1, essentially all $AH$ events are eliminated by
cut 1a (see Table~\ref{table:1}).  For channel 2, only the behavior at
large $\tan\beta$ is critical, and in this region, the $AH$
contribution is eliminated by $B(H\to W^+W^- ) \approx 0$.

Finally, we note that we have not included possible supersymmetric
backgrounds.  If present, such backgrounds certainly require more
study.  However, it is likely that slepton, neutralino, and chargino
pair production will be greatly reduced by our demands for multiple
$b$ tags and the accompanying cuts.  Bottom and top squark pair
production may be the leading SUSY backgrounds, but motivated by the
fact that such particles carry SU(3)$_{\text{C}}$ quantum numbers and
so are likely to be heavy, we also do not consider such processes in
our analysis.

\subsection{Systematic errors}
\label{sec:Systematics}

In this study, a number of systematic errors must be included.  Two
important sources are uncertainties in the $b$-tagging efficiency
$\epsilon_b$ and the running quark mass $m_b$.  In addition, however,
the determination of $\tan\beta$ may be degraded by uncertainties
arising from the virtual effects of other SUSY particles on Higgs
processes, which depend on unknown SUSY parameters. We will
incorporate all such uncertainties in our analysis as systematic
uncertainties, and in this section we describe them and give numerical
values for these errors. We note, however, that if other measurements
are available, these systematic uncertainties may be greatly reduced.
In Sec.~\ref{sec:Results}, we will consider the beneficial effects
that other measurements may have on our analysis.

As $b$-tagging plays a central role in our analysis, it is clear that
an accurate knowledge of the $b$-tagging efficiency is important.  We
have included a systematic uncertainty of $\pm 2\%$ for
$\epsilon_b$\cite{Jacksonconversation}.

The running $b$ quark mass enters the branching ratio formulas of
Eqs.~(\ref{bra}) and (\ref{brb}). As studied in Ref.~\cite{JLCI}, a
measurement of this parameter that is relatively free of theoretical
ambiguities from the $b\bar{b}$ potential and renormalization group
equation evaluations is possible using the branching ratio of $h$.
Ref.~\cite{JLCI} estimates that $B(h\to \tau\bar{\tau})$ may be
measured at a future $e^+e^-$ collider to $\sim 0.5\%$, leading to a
$1\sigma$ error on $m_b$ of 150 MeV, and we therefore take $m_b$ to be
in the range $m_b = 3.2 \pm 0.15$ GeV.

As noted in Sec.~\ref{subsec:Definitions}, we do not assume tree level
or specific 1--loop relations in the Higgs sector, but instead will
assume these are all independently measured.  The uncertainties in
these measurements then enter our analysis as systematic errors.  We
consider the Higgs masses and interactions in turn.  The $h$ mass will
be measured very precisely, and errors arising from this measurement
are negligible for this study.  The charged Higgs mass may be measured
through its $t\bar{b}\bar{t}b$ decay mode.  Ref.~\cite{Sop} finds that
given underlying masses of $m_{H^{\pm}}= 180$ GeV and $m_t= 140$ GeV,
the charged Higgs mass resolution is 16 GeV.  This measurement is
likely to be improved if supplemented by information from the
$tb\tau\nu$ decay mode.  Here, however, we adopt conservatively the
error given in Ref.~\cite{Sop} with appropriate rescaling, {\em i.e.},
we take $\delta m_{H^{\pm}}= 16\text{ GeV}/\sqrt{N}$, where $N$ is the
number of $t\bar{b}\bar{t}b$ charged Higgs events multiplied by the
efficiency of 3.5\% given in Ref.~\cite{Sop}.  Studies have not been
conducted for the $A$ and $H$ masses.  We will assume, however, that
their masses may be measured to the same accuracy as the charged
Higgs.

The parameter $\cos^2 (\beta-\alpha)$ may be determined by
$\sigma(e^+e^-\to Zh)$ to an accuracy of 2\%\cite{JLCI}, and we take
this as its systematic uncertainty. As given in Eq.~(\ref{Hhh}), there
may also be large radiative corrections to the $Hhh$ vertex.  In
Ref.~\cite{DHZ}, the measurement of this vertex through the branching
ratio $B(H\to hh)$ has been considered using the process $e^+e^- \to
\nu\bar{\nu} H$, followed by $H\to hh$.  The cross section $\sigma
(e^+e^-\to H\nu\bar{\nu})$ is suppressed by $\cos^2(\beta-\alpha )$,
but may be significant for low $\tan\beta$, the region in which an
accurate measurement of $B(H\to hh)$ is important for this study.  For
example, in Ref.~\cite{DHZ}, the cross section for this process with
$\tan\beta=1.5$, $m_H=200$ GeV, and $\sqrt{s}=500$ GeV is shown to be
3 fb, leading to hundreds of events per year.  Thus, when $B(H\to hh)$
is large enough to be important for this study, a fairly accurate
measurement of its value may be obtained.  Without detailed studies of
backgrounds, it is impossible to determine exactly what bounds may be
placed on $B(H\to hh)$; for this study, we simply estimate that
$B(H\to hh)$ may be measured with an error of 10\%, and include this
in our systematic errors.  It should be noted, however, that if the
$\nu\bar{\nu} H$ cross section is suppressed, one must turn to $AH$
production, and perform a global fit to $\lambda_{Hhh}$, $\tan\beta$
and possibly other parameters.  Such a fit is beyond our present
analysis.

Finally, there are radiative corrections to the decay widths and
production cross sections.  Those that depend on standard model
parameters are predictable, and so even if large may be incorporated
in the analysis, once calculated.  However, those that depend on
unknown SUSY parameters are more dangerous.  SUSY QCD corrections to
the hadronic decay width of the charged Higgs have been
calculated\cite{SUSYQCD}.  These studies have shown that the
corrections may in general be large and of order 40\%.  However, for
$200 \text{ GeV}\alt m_{H^{\pm}} \alt 400$ GeV and squark masses above
500 GeV, the SUSY QCD correction is reduced to 10 -- 20\%. SUSY QCD
corrections to neutral Higgs decay widths have been studied in
Ref.~\cite{JS}, with similar results.  We therefore include 20\%
systematic errors for the five decay widths $\Gamma(H^{\pm} \to tb)$,
$\Gamma(A \to b\bar{b})$, $\Gamma(A \to t\bar{t})$, $\Gamma(H \to
b\bar{b})$, and $\Gamma(H \to t\bar{t})$.

The electroweak corrections to the cross section come from diagrams
involving squarks, neutralinos, charginos, and sleptons.  Such
corrections have been studied for charged Higgs
production\cite{Gilbert}, where effects are found to be typically of
order 10\%, and may be as large as 25\%.  However, for a given range
of $\tan\beta$, the bounds will be determined primarily by channels in
which the cross sections scale as $\tan^2\beta$ or some higher power
of $\tan\beta$.  The uncertainty induced in $\tan\beta$ is then $\alt
5-12\%$, which will be seen to be negligible relative to other errors
in this study, and we therefore do not include this uncertainty.

\section{Results}
\label{sec:Results}

In this section we present quantitative results for the bounds on
$\tan\beta$ that may be achieved.  For now, we assume that SUSY decay
modes are absent --- such decay modes will be considered in
Sec.~\ref{sec:SUSYdecays}.  We first discuss how we include
statistical and systematic errors in our calculations of confidence
level contours, and then present results for a variety of underlying
parameters and experimental assumptions.

To bound $\tan\beta$, we must first select a set of underlying SUSY
parameters to determine the underlying physics scenario that we hope
to constrain.  In our framework, as discussed in Sec.~\ref{sec:Tanb},
this requires us to choose $\tan\beta$ and $m_{H^{\pm}}$, and also
$m_{\tilde{t}}$ to fix the radiative corrections.  This then fixes the
number of events that will be observed in each channel. (We assume for
simplicity that the number of observed events is given by the central
value corresponding to the underlying parameters.)

As described above, our analysis is general in that it does not assume
any fixed form of the Higgs radiative corrections.  Thus, to determine
$\tan\beta$ experimentally, we begin by taking the Higgs masses,
$\cos^2 (\beta-\alpha )$, and $B(H\to hh)$ to be bounded
experimentally through the methods described above.  Given these
measurements, the only remaining unknown parameter is $\tan\beta$.  To
determine $\tan\beta$, we postulate a hypothetical value $\tan\beta'$,
and determine if such a value is consistent with the observed numbers
of events in each of our eight channel.  To quantify this consistency,
we define a simple $\Delta \chi^2$ variable,

\begin{equation}
\Delta \chi^2 = \sum_{i=1}^8 \frac{(N_i - N'_i)^2}
{{\sigma^i}_{\text{stat}}^2 + {\sigma^i}_{\text{sys}}^2} \ ,
\end{equation}
where $i$ is summed over all channels, and $N_i$ ($N'_i$) is the
number of events in channel $i$ determined by the underlying
(postulated) value $\tan\beta$ ($\tan\beta'$).  The quantities
$\sigma_{\text{stat}}^i$ and $\sigma_{\text{sys}}^i$ are the
statistical and systematic errors for channel $i$, respectively, and
for simplicity, we add these in quadrature.  The statistical error is
$\sigma_{\text{stat}}^i = \sqrt{N'_i}$.  The systematic error is given
by

\begin{equation}
{\sigma^i}_{\text{sys}}^2 = \sum_{j=1}^{12} \left[ \frac{\partial
N'_i}{\partial P_j} \Delta P_j \right] ^2 \ ,
\end{equation}
where the sum is over the systematic uncertainties in the 12
quantities $P_j = \epsilon_b$, $m_b$, $m_{H^{\pm}}$, $m_A$, $m_H$,
$\cos^2 (\beta - \alpha )$, $B(H\to hh)$, $\Gamma(H^{\pm} \to tb)$,
$\Gamma(A \to b\bar{b})$, $\Gamma(A \to t\bar{t})$, $\Gamma(H \to
b\bar{b})$, and $\Gamma(H \to t\bar{t})$.  The deviations $\Delta P_j$
are the systematic uncertainties described for each quantity $P_j$ in
Sec.~\ref{sec:Systematics}.  In the following, we will display $\Delta
\chi^2 = 3.84$ contours in the $(\tan\beta, \tan\beta')$ plane, which
we will refer to as 95\% C.L. contours.

In these plots, the underlying scenario is determined by fixing
$\tan\beta$, $m_{H^{\pm}}$, and $m_{\tilde{t}}=1$ TeV.  Note that the
underlying value of $m_h$ therefore varies with $\tan\beta$.  Of
course, once $m_h$ is known, one should consider only scenarios that
predict $m_h$ in the experimentally allowed range. However, without
knowing $m_h$, we prefer to display results for scenarios with $m_h$
given by reasonable radiative corrections (in this case, radiative
corrections that may be produced by $m_{\tilde{t}}=1$ TeV).

\begin{figure}
\vspace*{.5in}
\centerline{
\psfig{file=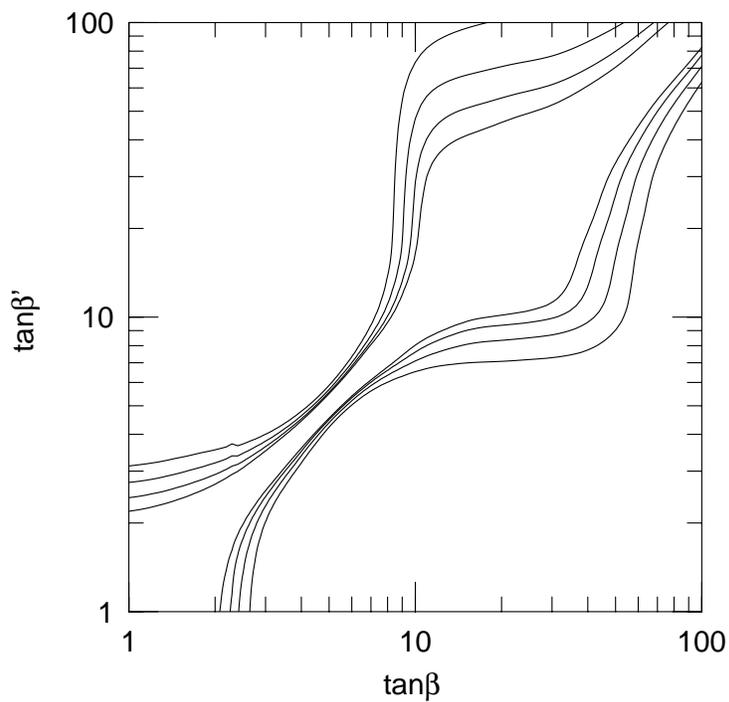,width=0.61\textwidth}
\vspace*{.5in}}
\caption{ 95\% C.L. bounds on $\tan\beta$ for $\protect\sqrt{s}=500$
GeV, $m_{H^{\pm}} = 200$ GeV, $\epsilon_b=60\%$, and four integrated
luminosities: 25, 50, 100, and 200 fb$^{-1}$ (from outside to inside).
For a fixed underlying value of $\tan\beta$, the values of
$\tan\beta'$ determined by the appropriate contours are the upper and
lower bounds that may be set experimentally.
\label{fig:result500}}
\end{figure}

We first consider the $\sqrt{s}=500$ GeV collider, and choose a
typical kinematically accessible charged Higgs mass of $m_{H^{\pm}} =
200$ GeV.  In Fig.~\ref{fig:result500}, we display 95\% C.L. contours
in the $(\tan\beta, \tan\beta')$ plane for four integrated
luminosities: 25, 50, 100, and 200 fb$^{-1}$, or 0.5, 1, 2, and 4
years at design luminosity.  For this plot, we assume $\epsilon_b =
60\%$, and have included all the systematic errors of
Sec.~\ref{sec:Systematics}.  We expect in this case to bound moderate
$\tan\beta$ stringently through the strong dependence of Higgs boson
branching fractions on $\tan\beta$ in this range.  In addition, we
expect to be able to bound large values of $\tan\beta$ through the
process $tbH^{\pm}$.  These characteristics are evident in
Fig.~\ref{fig:result500}.  As examples, we find that for an integrated
luminosity of 100 fb$^{-1}$ and the underlying values of $\tan\beta$
listed below, the 95\% C.L. bounds that may be obtained are

\begin{eqnarray}
\tan\beta = 2:  & \tan\beta' < 2.9 \ , \nonumber \\ 
\tan\beta = 3:  & 2.5 < \tan\beta' < 3.6 \ , \nonumber \\
\tan\beta = 5:  & 4.5 < \tan\beta' < 5.5 \ , \\
\tan\beta = 10: & 7.6 < \tan\beta' < 30 \ , \nonumber \\
\tan\beta = 60: & 40  < \tan\beta' < 90 \ . \nonumber 
\end{eqnarray}
Note that Yukawa coupling constants become non-perturbative below the
GUT scale if $\tan\beta$ is too close to 1, or if $\tan\beta$ is too
large ($\tan\beta\agt 70-80$). Thus, in much of the parameter space
that is theoretically interesting, significant constraints on
$\tan\beta$ may be obtained.

The above results have interesting implications as tests of Yukawa
coupling constant unification. For example, $m_b$--$m_\tau$
unification based on GUTs prefers either large ($\tan\beta\agt 60$) or
small ($\tan\beta\alt 2$) values of $\tan\beta$ \cite{b-tau}.
Furthermore, if we assume simple SO(10)-like unification\cite{so(10)},
$\tan\beta$ is approximately given by $m_t/m_b\approx 50-60$, since
the Yukawa couplings for top and bottom quarks are unified at the GUT
scale. As one can see in Fig.~\ref{fig:result500}, the values of
$\tan\beta$ predicted by these scenarios can be easily distinguished.
In addition, the stringent constraints on $\tan\beta$ available in its
moderate range are very useful for soft scalar mass determination, as
will be seen in Sec.~\ref{sec:Soft}.

\begin{figure}
\vspace*{.5in}
\centerline{
\psfig{file=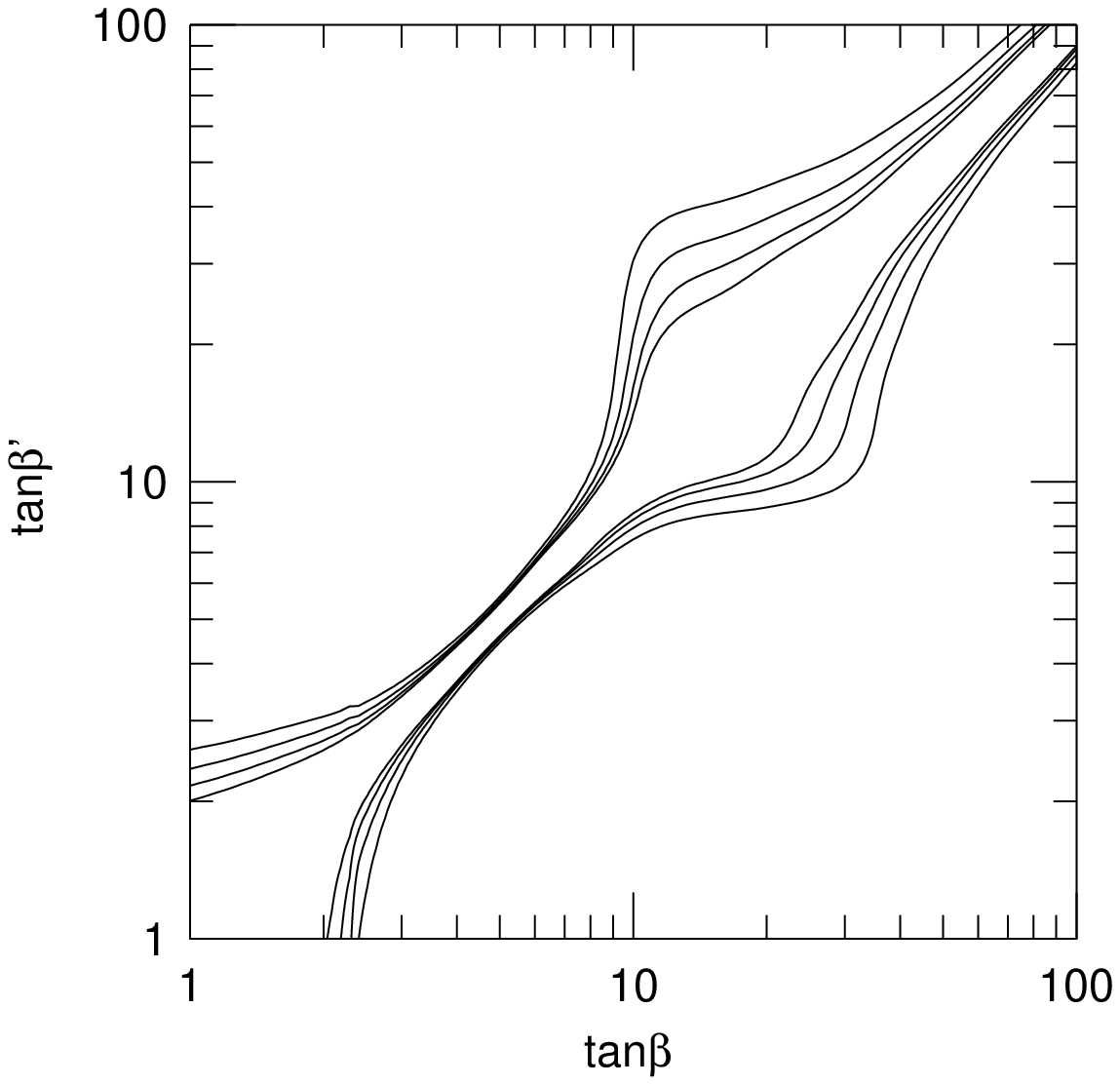,width=0.61\textwidth}
\vspace*{.5in}}
\caption{ 95\% C.L. bounds on $\tan\beta$ for $\protect\sqrt{s}=1$
TeV, $m_{H^{\pm}} = 200$ GeV, $\epsilon_b=60\%$, and four integrated
luminosities: 100, 200, 400, and 800 fb$^{-1}$.
\label{fig:result1000200}}
\end{figure}

\begin{figure}
\vspace*{.5in}\centerline{
\psfig{file=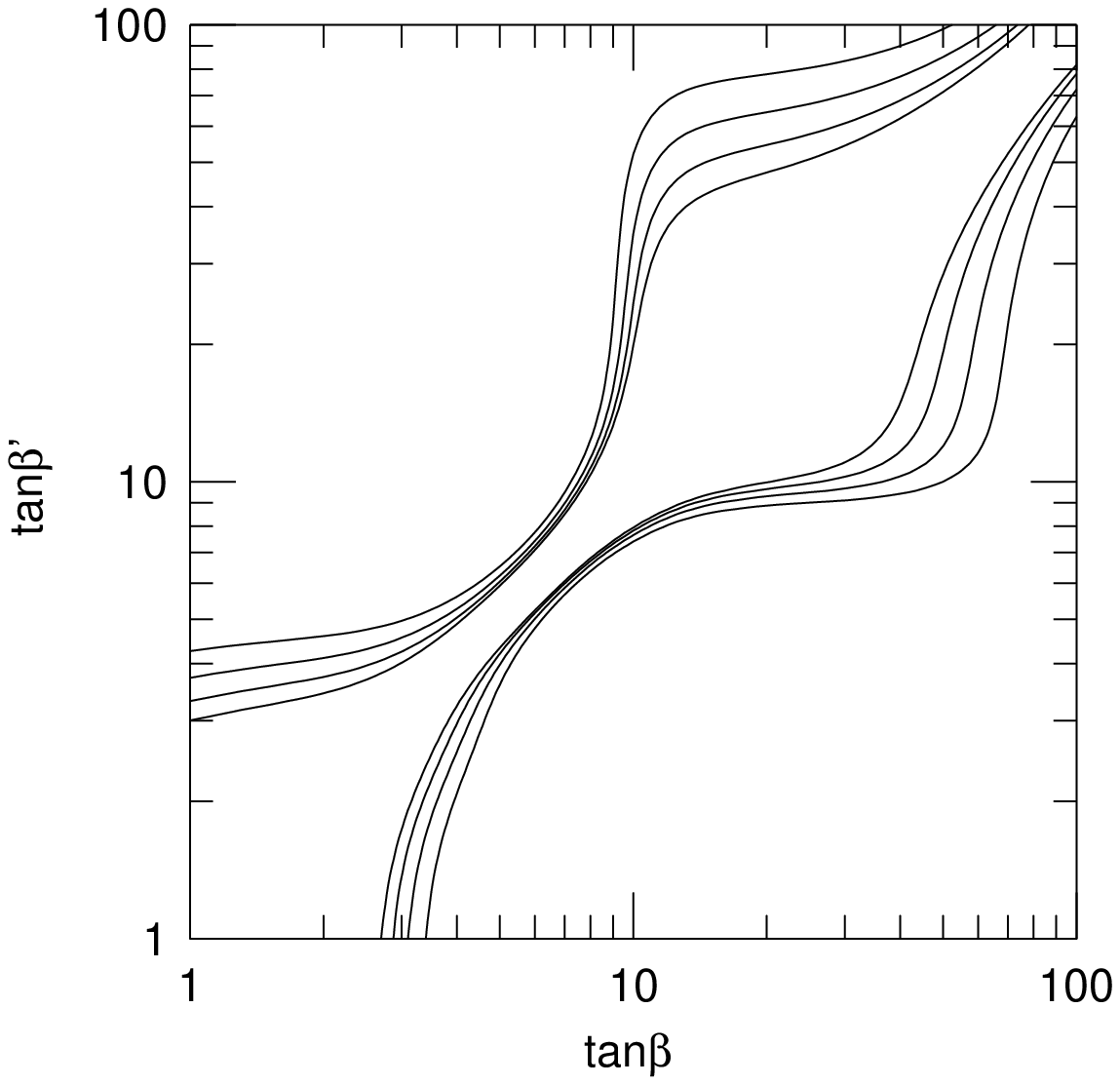,width=0.61\textwidth}
\vspace*{.5in}}
\caption{ 95\% C.L. bounds on $\tan\beta$ for $\protect\sqrt{s}=1$
TeV, $m_{H^{\pm}} = 300$ GeV, $\epsilon_b=60\%$, and four integrated
luminosities: 100, 200, 400, and 800 fb$^{-1}$.
\label{fig:result1000300}}
\end{figure}

\begin{figure}
\vspace*{.5in}\centerline{
\psfig{file=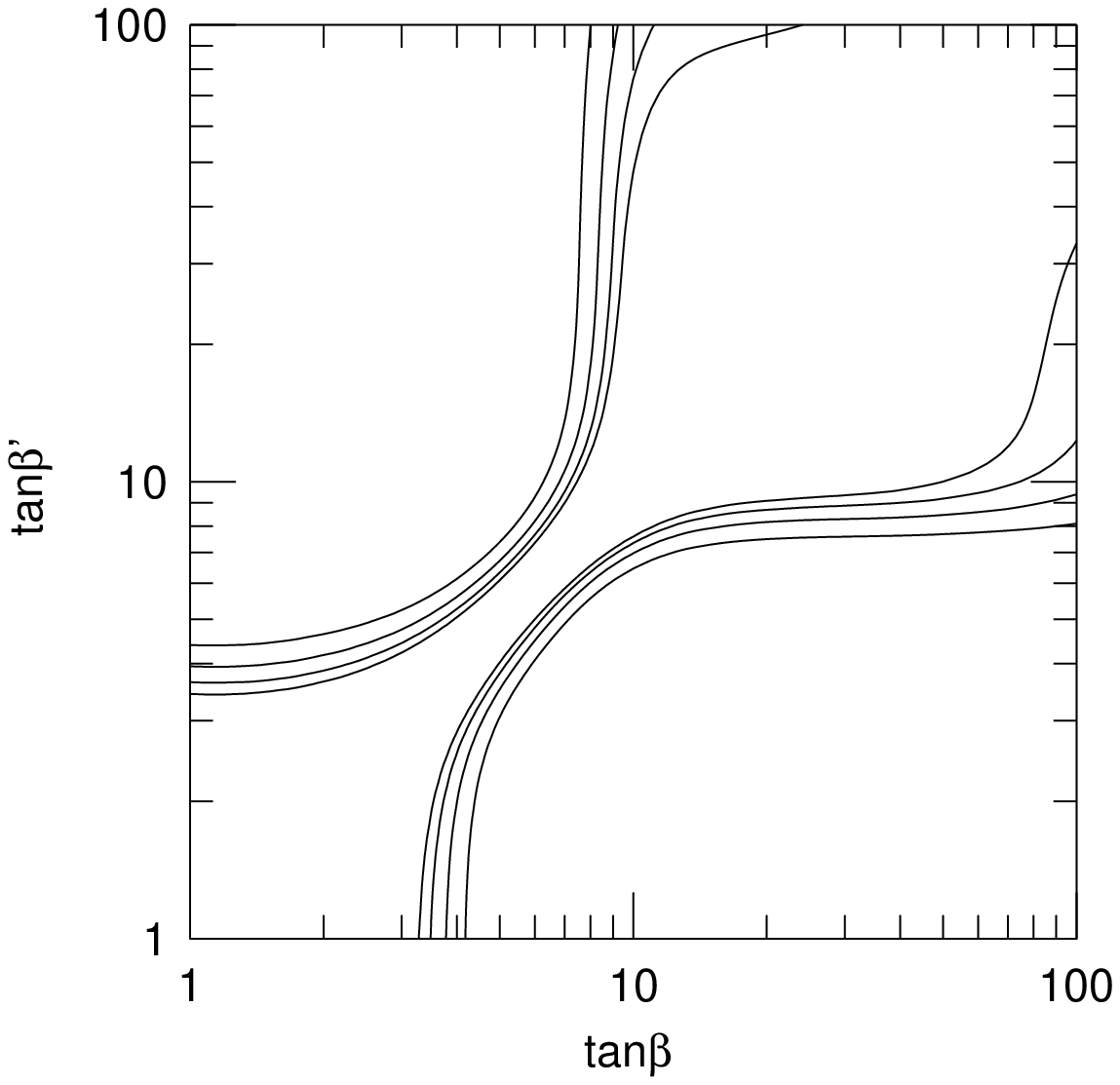,width=0.61\textwidth}
\vspace*{.5in}}
\caption{ 95\% C.L. bounds on $\tan\beta$ for $\protect\sqrt{s}=1$
TeV, $m_{H^{\pm}} = 400$ GeV, $\epsilon_b=60\%$, and four integrated
luminosities: 100, 200, 400, and 800 fb$^{-1}$.
\label{fig:result1000400}}
\end{figure}

If the heavy Higgs bosons are not produced in the first phase of a
future $e^+e^-$ collider's run, or even if they are, it may be
advantageous to increase the beam energy.  We consider next a
$\sqrt{s}=1$ TeV collider. In Figs.~\ref{fig:result1000200},
\ref{fig:result1000300}, and \ref{fig:result1000400}, we present
results for scenarios with this higher beam energy and $m_{H^{\pm}} =
200$, 300, and 400 GeV, respectively.  We plot contours for integrated
luminosities of 100, 200, 400, and 800 fb$^{-1}$.  For $m_{H^{\pm}} =
200$ GeV, the result is dramatically improved over the $\sqrt{s} =
500$ GeV case. This is in many ways a nearly ideal scenario for this
analysis.  As charged Higgs decays to $tb$ are still considerably
suppressed by phase space, $B(H^{\pm} \to \tau\nu)$ rises rapidly for
increasing $\tan\beta$.  This, in conjunction with the large
luminosities that are expected to be available at $\sqrt{s} = 1$ TeV,
implies that the statistical errors in channels 1 and 2 are greatly
reduced.  We see in this case that the bounds are stringent throughout
the range of $\tan\beta$, and, for example, for an integrated
luminosity of 100 fb$^{-1}$ and $\tan\beta = 60$, we may constrain
$\tan\beta'$ to the range $50 < \tan\beta' < 75$.

For $m_{H^{\pm}} = 300$ GeV, the bounds are slightly worse, as
$B(H^{\pm} \to tb)$ is now not highly suppressed by phase space, and
the number of events in channels 1 and 2 is therefore reduced.  In
addition, the power of channel 2 is reduced by the great increase in
$e^+e^- t\bar{t}$ background for $\sqrt{s} = 1$ TeV, relative to
$\sqrt{s}=500$ GeV. Improved cuts may be able to reduce this
background and improve the high $\tan\beta$ results.  Nevertheless,
the bounds are still quite strong for moderate $\tan\beta$, and
interesting determinations of high $\tan\beta$ are possible for large
integrated luminosities.  Finally, for $m_{H^{\pm}} = 400$ GeV, the
bounds are again weaker, but we are still able to distinguish low,
moderate, and high $\tan\beta$, and the measurement will be useful for
soft scalar mass determinations, as we will see below.

\begin{figure}
\vspace*{.5in}\centerline{
\psfig{file=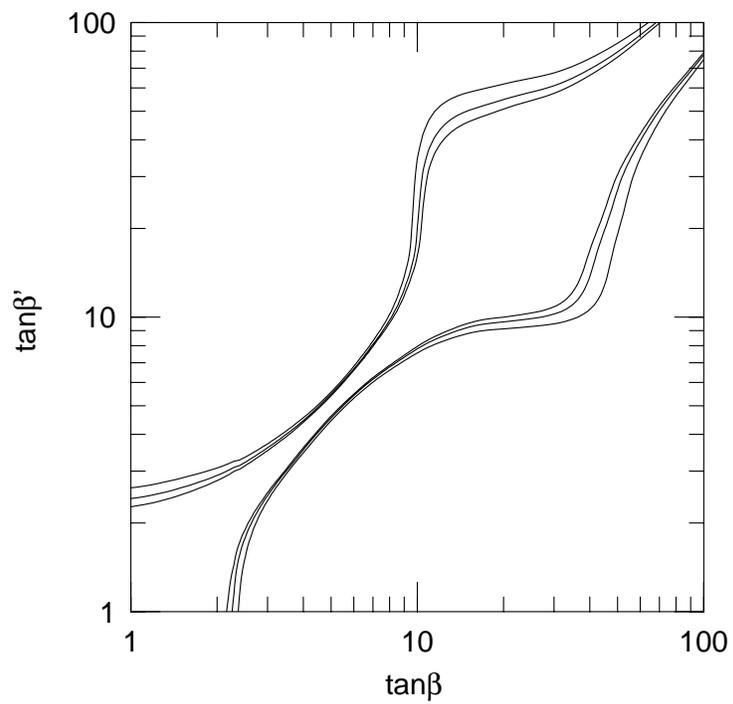,width=0.61\textwidth}
\vspace*{.5in}}
\caption{ 95\% C.L. bounds on $\tan\beta$ for $\protect\sqrt{s}=500$
GeV, $m_{H^{\pm}} = 200$ GeV, an integrated luminosity of 100
fb$^{-1}$, and $\epsilon_b=50$, 60, and 70\% (from outside to inside).
\label{fig:resulteb}}
\end{figure}

We next consider the dependence of our results on the assumed
$b$-tagging efficiency.  In Fig.~\ref{fig:resulteb}, we again consider
the case $\sqrt{s}=500$ GeV and $m_{H^{\pm}} = 200$ GeV, but plot
bounds for a fixed luminosity of 100 fb$^{-1}$ and three $b$-tagging
efficiencies $\epsilon_b = 50\%$, 60\%, and 70\%.  We see that the
effect of increased $\epsilon_b$ is roughly to decrease the integrated
luminosity required to achieve a certain bound.

We conclude this section with a discussion of the leading sources of
systematic errors in the results displayed in
Figs.~\ref{fig:result500} -- \ref{fig:result1000400}. As can be seen
in the figures, the weakest bounds are achieved in the small
$\tan\beta$ region ($\tan\beta \alt 2-3$) and the large $\tan\beta$
region.  In the low $\tan\beta$ region, and for the $m_{H^\pm}=200$
and 300 GeV cases, $\tan\beta$ is mainly constrained by channel 1 (the
``$H^+H^-$'' channel).  The reason for this is that, although channels
3 -- 8 are sensitive to variations in $\tan\beta$, as may be seen in
Figs.~\ref{fig:channels1} -- \ref{fig:channels4}, these channels
require many $b$ tags.  The uncertainty in $\epsilon_b$ thus
significantly weakens the bounds from these channels and is, in fact,
the leading systematic error.  However, if the systematic uncertainty
in $\epsilon_b$ is reduced, these channels may improve the
constraints.  For $m_{H^\pm}=400$ GeV, channel 1 loses its
significance since $B(H^\pm\rightarrow\tau\nu )$ is highly suppressed,
and channel 3, which has the largest cross section among the inportant
processes, becomes the most sensitive one to $\tan\beta$. Again, the
largest systematic error is the uncertainty in $\epsilon_b$.  Thus, in
the low $\tan\beta$ region, for all $m_{H^\pm}$ considered, the
results can be improved if we can reduce the uncertainty in
$\epsilon_b$, though the statistical error is also non-negligible.

\begin{figure}
\vspace*{.5in}\centerline{
\psfig{file=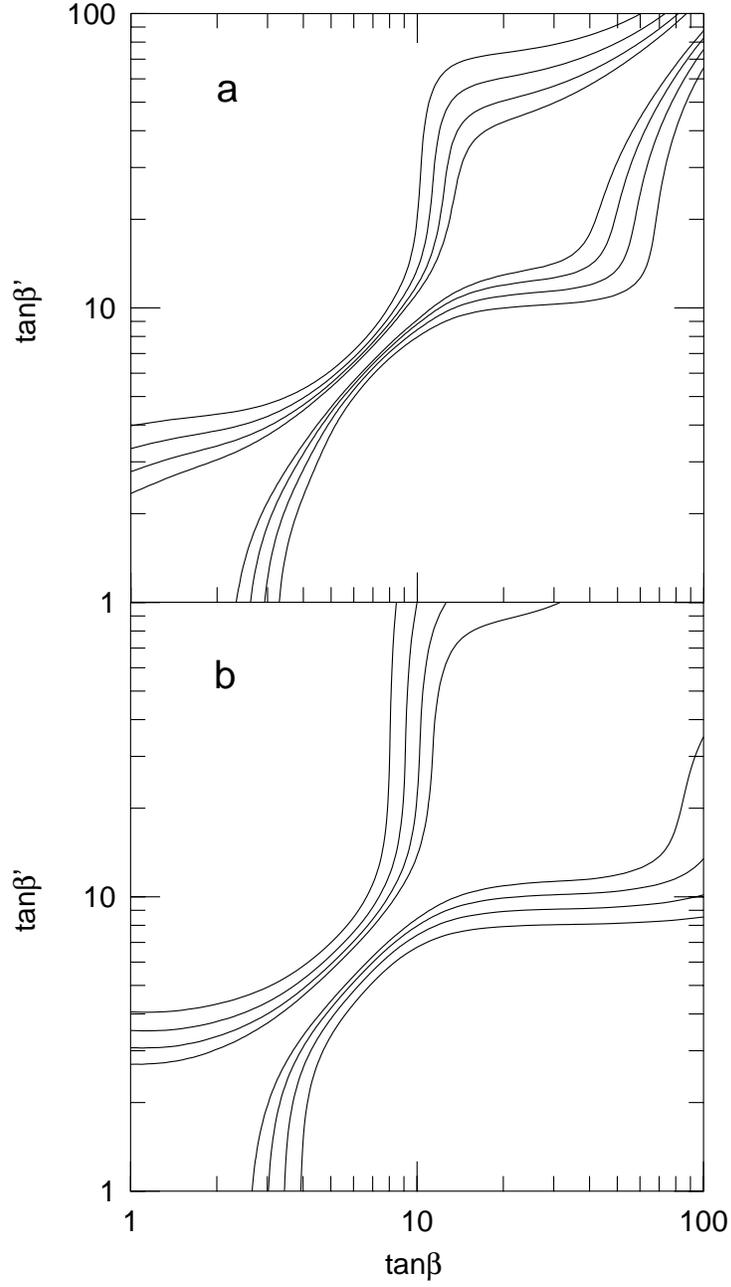,width=0.61\textwidth}
\vspace*{.5in}}
\caption{ 95\% C.L. bounds on $\tan\beta$ for (a) $m_{H^{\pm}} =300$
GeV and (b) $m_{H^{\pm}} = 400$ GeV, $\protect\sqrt{s}=1$ TeV,
$\epsilon_b=60\%$, and four integrated luminosities: 100, 200, 400 and
800 fb$^{-1}$ with all systematic uncertainties omitted.
\label{fig:result1000300_nosys}}
\end{figure}

For large $\tan\beta$, $\tan\beta$ is constrained only by channel 2
(the ``$tbH^\pm$'' process). In this case, the primary source of error
is statistical, and for typical luminosities, a reduction of the
systematic errors does not substantially improve the result. However,
for $m_{H^\pm}=200\text{ GeV}$ or 300 GeV and large $\tan\beta$, the
systematic uncertainty is not negligible if a high luminosity is
obtained.  In this case, the leading sources of systematic error are
$m_{H^{\pm}}$, $m_b$ and $\Gamma(H^{\pm} \to tb)$, and the results may
be noticeably improved if these errors are reduced. In
Fig.~\ref{fig:result1000300_nosys}, we show the 95\% C.L. contours
with all systematic uncertainties omitted for $m_{H^{\pm}} = 300$ GeV
and 400 GeV.  Comparing these figures with
Figs.~\ref{fig:result1000300} and \ref{fig:result1000400}, we see that
the results may be improved significantly if the systematic errors are
greatly reduced.

Throughout this analysis, we have assumed that no detailed knowledge
of the radiative corrections to the Higgs sector may be obtained, and
we therefore rely on experimental measurements of the various Higgs
masses and couplings.  However, if these corrections are
well-understood, for example, through detailed measurements of top
squark masses and left-right mixing, the results of this analysis may
be improved significantly.  For example, if the radiative corrections
are highly constrained, the triple Higgs vertex is essentially a
function of $\tan\beta$ only, and we need not rely on a measurement of
$B(H\to hh)$.  There is then a strong dependence of the multi-$b$
cross sections on $\tan\beta$.  We have analyzed this possibility, and
find, in particular, that channel 8 is then strongly dependent on
$\tan\beta$, and this leads to marked improvements in the low
$\tan\beta$ region.  This is but one example of how information from
other sectors may improve these results.  It is clear that other
measurements from the LHC or NLC may significantly improve the results
presented here.

\section{Scenarios with SUSY Decay Modes}
\label{sec:SUSYdecays}

Up to this point, we have assumed that Higgs scalars decay only to
standard model particles.  For large Higgs masses, however, decays to
supersymmetric particles may be allowed
\cite{DKZchargino,Squarkdecays}.  In this section, we discuss the
effects that decays to sleptons, neutralinos, and charginos have on
our analysis.  Squarks are typically heavy, and so decays to them will
not be considered here.

In many models, the right-handed charged sleptons are the lightest
sfermions, as their masses are not increased by SU(3)$_{\text{C}}$ or
SU(2)$_{\text{L}}$ interactions in the renormalization to low
energies.  We will therefore begin by considering the scenario in
which heavy Higgs decays to pairs $\tilde{l}_R^* \tilde{l}_R$ are open
and all other SUSY decays are closed.  The scalars $A$ and $H^{\pm}$
may decay only to $\tilde{l}_R^* \tilde{l}_L$, and so their branching
fractions are unchanged in the absence of $\tilde{l}_R$--$\tilde{l}_L$
mixing. On the other hand, the $H$ boson may decay into $\tilde{l}_R^*
\tilde{l}_R$ pairs through a $D$-term interaction.  The important
point is that the $H\tilde{l}_R^* \tilde{l}_R$ vertex is completely
fixed by the U(1)$_{\rm Y}$ gauge coupling constant and $\tan\beta$,
{\em i.e.}, it does not depend on additional unknown SUSY parameters.
In addition, the slepton masses will be very accurately measured at
future $e^+e^-$ colliders\cite{JLC,BV}.  Therefore, in the case where
$H\rightarrow\tilde{l}_R^* \tilde{l}_R$ is the only relevant SUSY
decay mode, no new systematic uncertainties enter our analysis. The
primary effect of this decay, then, is only to decrease the number of
the signal events from $AH$ production.

\begin{figure}
\vspace*{.5in}\centerline{
\psfig{file=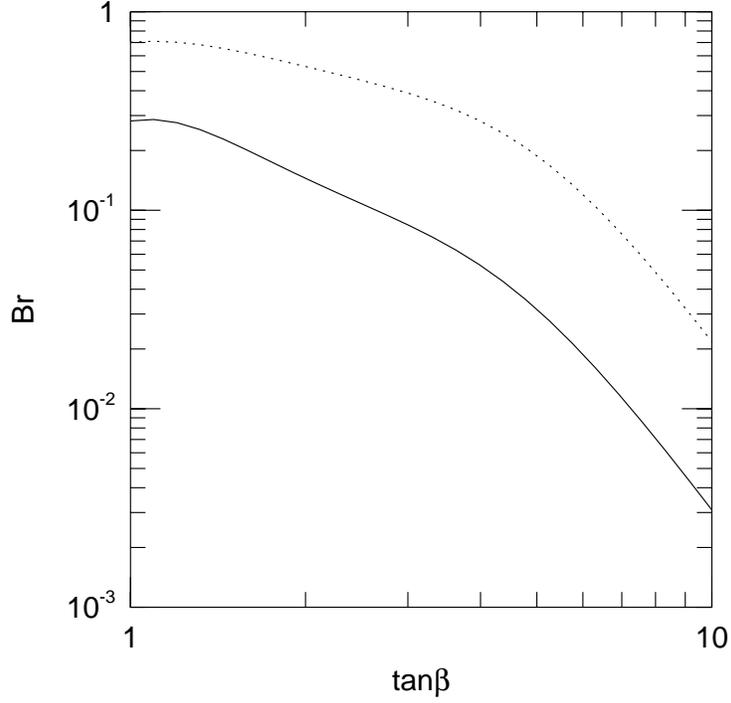,width=0.61\textwidth}
\vspace*{.5in}}
\caption{ Branching ratios $B(H \to \tilde{l}_R^* \tilde{l}_R)$
(solid) and $B(H \to \tilde{l}_L^* \tilde{l}_L) + B(H\to
\tilde{\nu}_L^* \tilde{\nu}_L)$ (dashed).  The $H$ mass and couplings
are determined for fixed $m_{H^{\pm}} = 300$ GeV, and include the
radiative correction of Eqs.~(\protect\ref{deltamassmatrix}) and
(\protect\ref{epsilon}) with $m_{\tilde{t}}= 1$ TeV.  For the solid
(dashed) curve, the three generations of right- (left-) handed
sleptons are assumed degenerate with mass $m_{\tilde{l}_R} = 100$ GeV
($m_{\tilde{l}_L} = 100$ GeV), and all other sparticles decay modes
are assumed closed.
\label{fig:sleptonLorR}}
\end{figure}

In Fig.~\ref{fig:sleptonLorR} the branching ratio of $H \to
\tilde{l}_R^* \tilde{l}_R$ is given by the solid curve for fixed
$m_{H^{\pm}} = 300$ GeV and three degenerate generations of
right-handed sleptons with masses $m_{\tilde{l}_R} = 100$ GeV.  We see
that the branching fraction never exceeds 0.3 and decreases rapidly
for increasing $\tan\beta$ as the width to $b$ quarks becomes
dominant.  In particular, for $\tan\beta\agt 3$, the range in which
our analysis may give stringent bounds, the branching ratio is less
than 0.1. Furthermore, if the decay mode $H \to t\bar{t}$ is open, the
branching ratio for this SUSY decay mode is suppressed even for the
low $\tan\beta$ region.

\begin{figure}
\vspace*{.5in}\centerline{
\psfig{file=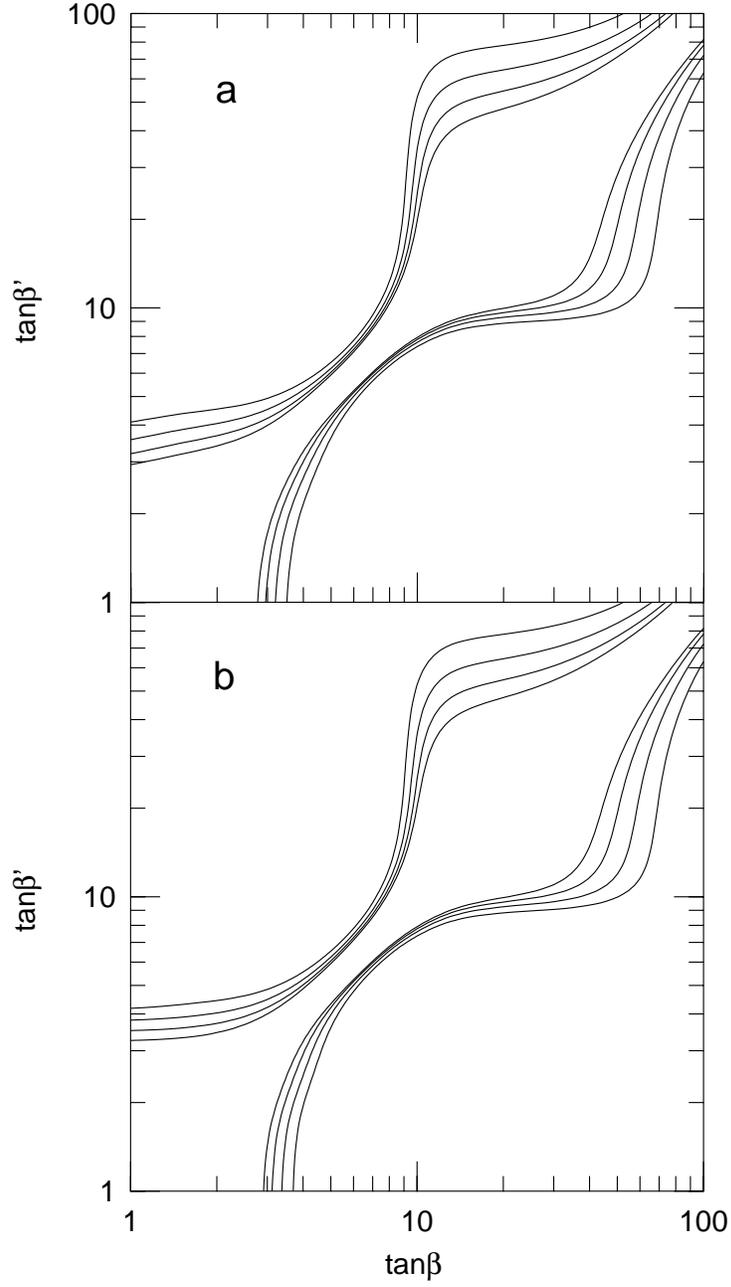,width=0.61\textwidth}
\vspace*{.5in}}
\caption{ 95\% C.L. bounds on $\tan\beta$ for $\protect\sqrt{s}=1$ TeV,
$m_{H^{\pm}} = 300$ GeV, $\epsilon_b=60\%$, and four integrated
luminosities: 100, 200, 400, and 800 fb$^{-1}$, with (a) only the
right-handed slepton decays open with $m_{\tilde{l}_R} = 100$, and (b)
only the left-handed slepton decays open with $m_{\tilde{l}_L} = 100$
GeV.  All systematic errors are included.
\label{fig:resultRL}}
\end{figure}

If only decays to left-handed slepton pairs are possible, again only
the $H$ branching fractions are affected.  In
Fig.~\ref{fig:sleptonLorR}, the solid curve gives the branching
fraction $B(H \to \tilde{l}_L^* \tilde{l}_L) + B(H\to \tilde{\nu}_L^*
\tilde{\nu}_L)$, again for fixed $m_{H^{\pm}} = 300$ GeV and assuming
three degenerate generations with masses $m_{\tilde{l}_L} = 100$ GeV
and $m_{\tilde{\nu}_L}$ determined by the relevant relations of
Eq.~(\ref{scalarmass}), which is given in the next section.  We see
that the branching ratio is enhanced relative to the previous case, as
the SU(2)$_{\text{L}}$ gauge couplings now contribute to the decay
process, and decays to both charged sleptons and sneutrinos are now
open.  However, the branching ratios again drop rapidly for increasing
$\tan\beta$.  In Fig.~\ref{fig:resultRL}, we give results for
$\tan\beta$ bounds with $\sqrt{s}=1$ TeV, $m_{H^{\pm}} = 300$ GeV, and
all systematic errors included, and including the effects of SUSY
decays to (a) right-handed sleptons with $m_{\tilde{l}_R} = 100$ GeV,
and (b) left-handed sleptons with $m_{\tilde{l}_L} = 100$ GeV. As one
can see, the loss in statistics is not very significant in both cases,
and Fig.~\ref{fig:result1000300} is almost unchanged.

If decays to both left- and right-handed sleptons are allowed, the
branching ratios of $H^{\pm}$, $A$, and $H$ are all altered.  Decays
to left-right pairs involve the $\mu$ parameter, as well as the
trilinear scalar couplings.  If these parameters are not measured,
they may contribute large systematic errors to the measurement of
$\tan\beta$.  Of course, these parameters may also be measured in
different processes, for example, from chargino and neutralino masses
for $\mu$ and left-right mixings for the trilinear scalar coupling.  A
complete analysis would therefore require a simultaneous fit to all of
these parameters.

Finally, we briefly consider decays to charginos and neutralinos.
These decays have been considered in detail \cite{DKZchargino}, and
have been shown to be dominant in some regions of parameter space.
However, if only decays to the lighter two neutralinos and the lighter
chargino are available, and these are either all gaugino-like or all
Higgsino-like, as is often the case, these decays are suppressed by
mixing angles.  If we are in the mixed region, these decay rates may
be large, but in this case, all six charginos and neutralinos should
be produced, and the phenomenology is quite rich and complicated.

\section{Determining Soft Scalar Masses}
\label{sec:Soft}

As mentioned in Sec.~\ref{sec:Introduction}, the parameter $\tan\beta$
plays an important role in determining the masses and interactions of
many supersymmetric particles.  A measurement of $\tan\beta$ from the
Higgs scalar sector is therefore valuable for constraining other
supersymmetric parameters of the theory, or for testing SUSY relations
in another sector.  In this section, we present one simple example,
namely, the determination of soft SUSY breaking scalar masses.

The physical masses of sleptons and squarks are given (in first
generation notation) by

\begin{equation}
\label{scalarmass}
\begin{array}{rcl}
m_{\tilde{u}_L}^2 &=&
m_{\tilde{Q}}^2+m_u^2+m_Z^2(\frac{1}{2}-\frac{2}{3}\sin^2 \theta_W )\cos
2\beta \ , \\ m_{\tilde{d}_L}^2 &=&
m_{\tilde{Q}}^2+m_d^2+m_Z^2(-\frac{1}{2}+\frac{1}{3}\sin^2 \theta_W
)\cos 2\beta \ , \\ m_{\tilde{u}_R}^2 &=&
m_{\tilde{U}}^2+m_u^2+m_Z^2(\frac{2}{3}\sin^2 \theta_W )\cos 2\beta \ , \\
m_{\tilde{d}_R}^2 &=& m_{\tilde{D}}^2+m_d^2+m_Z^2(-\frac{1}{3}\sin^2
\theta_W )\cos 2\beta \ , \\ m_{\tilde{e}_L}^2 &=&
m_{\tilde{L}}^2+m_e^2+m_Z^2(-\frac{1}{2}+\sin^2 \theta_W )\cos 2\beta \ , \\
m_{\tilde{\nu}_L}^2 &=& m_{\tilde{L}}^2+m_Z^2(\frac{1}{2})\cos 2\beta \ , \\
m_{\tilde{e}_R}^2 &=& m_{\tilde{E}}^2+m_e^2+m_Z^2(-\sin^2 \theta_W )\cos
2\beta \ ,
\end{array}
\end{equation}
where $m_{\tilde{Q}}$, $m_{\tilde{U}}$, $m_{\tilde{D}}$,
$m_{\tilde{L}}$, and $m_{\tilde{E}}$ are the soft SUSY breaking scalar
masses.  In these relations, possible mixings among sfermions are
neglected.  In fact, such mixings may be large and lead to a variety
of new phenomena that may be probed at future $e^+e^-$ colliders.
Left-right mixing, which may be large for third generation sfermions,
has been analyzed for scalar taus\cite{Nojiri}, and intergenerational
slepton mixing has also been studied recently\cite{ACFH}. For
simplicity, however, we assume in this section that these effects are
absent.

As emphasized in Ref.~\cite{MEP}, the pattern of soft SUSY breaking
parameters is a reflection of the SUSY breaking mechanism, and so
accurate determinations of the soft SUSY breaking masses may provide
insights into the physics of SUSY breaking.  In addition, accurate
measurements of the sfermion masses may help determine the gauge
and/or flavor structures at higher energies
\cite{scalar_mass1,scalar_mass2}. As can be seen in
Eq.~(\ref{scalarmass}), an accurate measurement of the soft scalar
masses requires precise measurements of both the physical sfermion
masses and $\tan\beta$.  If $\tan\beta$ is completely unknown, the
uncertainty in the soft scalar mass is considerably greater than the
expected uncertainty from measurements of the physical masses.  For
example, if $m_{\tilde{e}_R} = 100$ GeV, the variation in
$m_{\tilde{E}}$ for $1 < \tan\beta < 60$ is 10 GeV.  On the other
hand, slepton and squark masses may be measured at $e^+e^-$ colliders
without parameter unification assumptions with a fractional error of
1--2\%\cite{JLC,BV,FF}.  One might hope, therefore, that a measurement
of $\tan\beta$ from the Higgs sector would reduce the uncertainty from
$\tan\beta$ to a comparable level.

\begin{figure}
\vspace*{.5in}\centerline{
\psfig{file=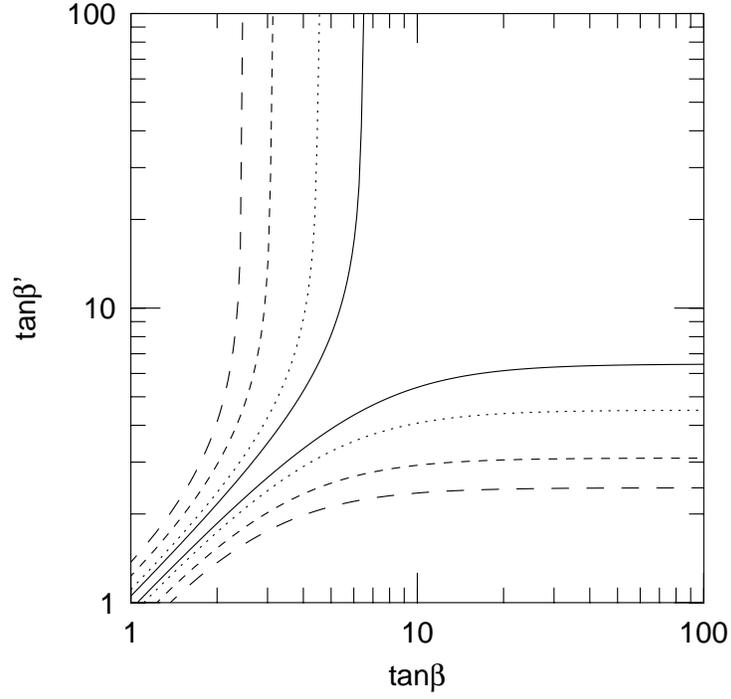,width=0.61\textwidth}
\vspace*{.5in}}
\caption{ Contours of constant $\Delta m =$ 0.5 GeV (solid), 1 GeV
(dotted), 2 GeV (short-dashed), and 3 GeV (long-dashed), where $\Delta
m = \left| m_{\tilde{E}} (\tan\beta ) - m_{\tilde{E}} (\tan\beta')
\right|$, for fixed physical mass $m_{\tilde{e}_R}= 100$ GeV.
\label{fig:scalarmass}}
\end{figure}

In Fig.~\ref{fig:scalarmass}, we plot contours of constant 

\begin{equation}
\Delta m = \left| m_{\tilde{E}} (\tan\beta ) - m_{\tilde{E}} (\tan\beta') 
\right| \ ,
\end{equation}
for fixed physical mass $m_{\tilde{e}_R}= 100$ GeV. For other sfermion
species and masses, the contour labels (0.5, 1, 2, and 3 GeV) should
be multiplied by approximately $(|F|/ \sin^2 \theta_W ) (100 \text{
GeV} / m_{\tilde{f}})$, where $F$ is the appropriate function of
$\sin^2 \theta_W$ in parentheses in Eq.~(\ref{scalarmass}). (For
example, for the right-handed slepton, $F=-\sin^2\theta_W$.) The soft
mass depends on $\tan\beta$ only through $\cos 2\beta$.  The
dependence is therefore very slight for large $\tan\beta$, and, as may
be seen in Fig.~\ref{fig:scalarmass}, a bound such as $\tan\beta \agt
6$ is already very powerful for the purposes of determining soft mass
parameters.  Comparing this with the bounds that may be achieved from
Higgs scalars, we see that for $\tan\beta \agt 3$, the uncertainty in
$m_{\tilde{E}}$ from $\tan\beta$ is reduced below that from the
physical mass measurement.

An independent measurement of $\tan\beta$ may also be possible in the
scalar sector if the masses of both sfermions of a left-handed doublet
are measured.  (Such a measurement is not necessarily easy, even if
sfermions are kinematically accessible --- in the slepton sector,
sneutrinos may decay invisibly; in the squark sector, such a
measurement requires a determination of quark flavor.) In this case, a
comparison of the two $\tan\beta$ determinations constitutes a highly
model-independent test of SUSY\cite{FMPT}, without any assumptions of
GUT or SUGRA relations.

\section{Implications for Top Squarks}
\label{sec:Upper}

In this section, we discuss another application of the $\tan\beta$
measurement, namely, the application to the top squark sector.  Top
squarks are singled out by the large top Yukawa coupling, which
implies that significant left-right stop mixing is generic, and that
radiative corrections from top-stop loops are highly significant in
determining the properties of the Higgs bosons.  For these reasons,
precise measurements of the parameters in the Higgs sector may allow
us to constrain parameters in the stop sector.

In the absence of left-right stop mixing, the leading radiative
corrections to the CP--even Higgs sector were given in
Eq.~(\ref{epsilon}).  In general, however, all parameters of the top
squark sector enter.  The top squark mass matrix is

\begin{equation}
\label{sup}
{\cal M}_{\tilde{t}}^2 =
 \left( \begin{array}{cc}
m_{\tilde{Q}}^2+m_t^2+m_Z^2(\frac{1}{2}-\frac{2}{3} \sin^2\theta_W )
\cos 2\beta &m_t(A_t-\mu\cot\beta)  \\
m_t(A_t-\mu\cot\beta)
&m_{\tilde{T}}^2+m_t^2+m_Z^2(\frac{2}{3} \sin^2 \theta_W )\cos 2\beta
\end{array} \right) \ ,
\end{equation}
where $m_{\tilde{Q}}$ and $m_{\tilde{T}}$ are soft SUSY breaking
parameters discussed in the previous section, $\mu$ is the
supersymmetric Higgs mass parameter given in Eq.~(\ref{Superpot}), and
$A_t$ is the top trilinear scalar coupling. The physical top squark
masses are the eigenvalues of this matrix, and we denote the lighter
and heavier top squarks as $\tilde{t}_1$ and $\tilde{t}_2$,
respectively.  For low and moderate $\tan\beta$, where the bottom
squark contributions may be neglected, the CP--even Higgs masses are
then determined by $m_A$ and $\tan\beta$ at tree level, and
$m_{\tilde{Q}}$, $m_{\tilde{T}}$, $\mu$, and $A_t$, all of which enter
through radiative corrections.

\begin{figure}
\vspace*{.5in}\centerline{
\psfig{file=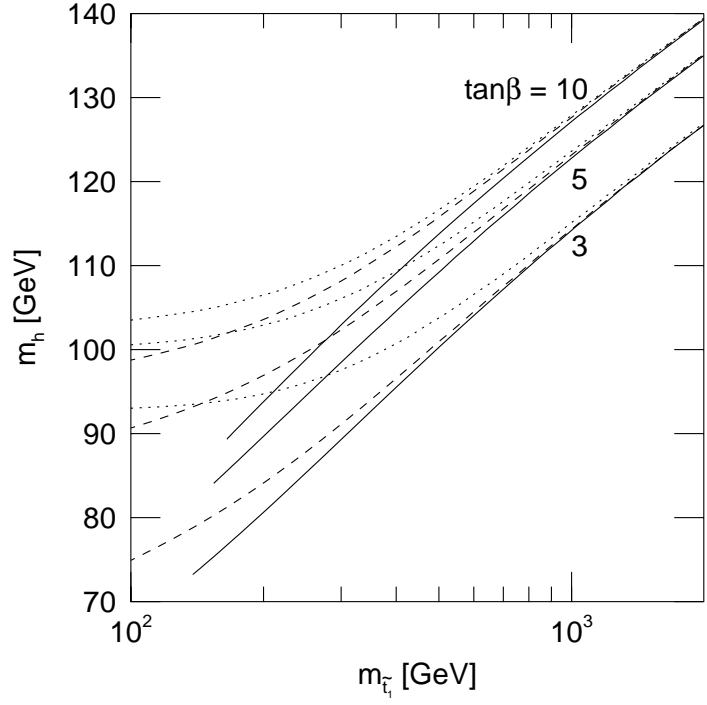,width=0.61\textwidth}
\vspace*{.5in}}
\caption{ The lightest Higgs mass $m_h$ as a function of the mass of
the lighter stop $m_{\tilde{t}_1}$ for $\tan\beta =3$, 5, and 10.
Here, we fix $\mu =200 \text{ GeV}$, $m_A = 300$ GeV, and $A_t=0$
(solid lines), $A_t=200\text{ GeV}$ (dashed lines), and $A_t=-200
\text{ GeV}$ (dotted lines).
\label{fig:stop}}
\end{figure}

In Fig.~\ref{fig:stop}, we plot $m_h$ as a function of
$m_{\tilde{t}_1}$ for various values of $\tan\beta$ and $A_t$.  Here,
we fix $\mu=200$ GeV and $m_A=300$ GeV, and we take $m_{\tilde{Q}} =
m_{\tilde{T}}$ for simplicity.  We see that there is strong dependence
of $m_h$ on the various top squark parameters, implying that we may be
able to constrain new parameters with a measurement of $\tan\beta$.
For example, if $\mu$ is measured from the gaugino-Higgsino sector, a
measurement of $\tan\beta$, along with measurements of $m_A$,
$m_{\tilde{t}_1}$ and $m_h$, may allow us to constrain $A_t$, a
parameter that otherwise may be rather difficult to measure without
model-dependent assumptions.  (Note that we have assumed the relation
$m_{\tilde{Q}} = m_{\tilde{T}}$ for simplicity.  This relation may be
relaxed if we also measure $m_{\tilde{t}_2}$ and impose this as an
additional constraint.)  On the other hand, from the figure, we also
see an asymptotic behavior for large $m_{\tilde{t}_1}$ --- if the soft
SUSY breaking parameters $m_{\tilde{Q}}$ and $m_{\tilde{T}}$ dominate
the left-right mixing terms, $m_h$ is simply a function of
$\tan\beta$, $m_A$, and $m_{\tilde{t}_1}$.  Thus, assuming that this
holds, even if top squarks are too heavy to be discovered at either a
future $e^+e^-$ collider or the LHC, we may be able to place {\em
upper} bounds on their masses by measuring $m_h$, $m_A$, and
$\tan\beta$.  It is important to note, however, that if $A_t$ and
$\mu$ may be arbitrarily large, one cannot draw such a conclusion.

In this section, we have not considered quantitatively the results
that may be achieved.  Clearly, measurements of many parameters enter
the analyses suggested here, and an overall fit to the relevant
parameters will be necessary in a complete analysis.  However, the
example of the top squark sector illustrates at least qualitatively
the possibility of applying an accurate $\tan\beta$ measurement to
interesting determinations of parameters in other sectors of
supersymmetric models.

\section{Conclusions}
\label{sec:Conclusions}

In this study we have considered the prospects for measuring
$\tan\beta$ through heavy Higgs scalar production and decay at a
future $e^+e^-$ collider.  The branching ratios of heavy Higgs scalars
are strongly dependent on $\tan\beta$.  In addition, we have seen that
Higgs signals typically have many $b$ quarks in the final state,
which, given the excellent $b$-tagging efficiencies and purities
expected, allows them to be separated from SM backgrounds in a number
of different channels.  The cross sections from these channels allow
us to significantly constrain the parameter space, and in particular,
to place bounds on $\tan\beta$.

We have relied on experimental measurements wherever possible.  The
neutral Higgs sector is subject to large radiative corrections,
depending strongly, for example, on top squark masses and mixings.  In
our analysis, we have not assumed that such radiative corrections are
small.  Instead, we treat the Higgs scalar masses, the parameter
$\cos^2(\beta - \alpha )$, and the vertex $\lambda_{Hhh}$ as
independent quantities, constrained only by experimental measurements.
In addition, we have avoided assumptions of SUSY parameter
unification.  The analysis method is therefore formally applicable to
models with arbitrary radiative corrections.  We have, however,
assumed a minimal Higgs sector throughout this analysis.  If
additional Higgs fields are present, determinations of $\tan\beta$
from the various channels may not be consistent, and the effects of a
non-minimal Higgs sector may also be detected by an analysis similar
to this one\cite{fm}.

We have considered scenarios with a variety of beam energies, Higgs
masses, integrated luminosities, $b$-tagging efficiencies, and
assumptions about systematic errors, and have also considered the
impact of SUSY decay modes.  For all scenarios considered, we find
that the strong dependence of heavy Higgs branching ratios on
$\tan\beta$ allows stringent constraints for moderate $\tan\beta$.
These results imply that for $\tan\beta\agt 3$, the soft scalar mass
parameter determinations are likely to be limited by the precision of
the corresponding physical scalar mass measurements.  In addition, we
have seen that the three body cross section $tbH^{\pm}$ grows rapidly
for large $\tan\beta$, and is large enough for some scenarios to
provide interesting constraints for large $\tan\beta$. These bounds
allow one to confirm or exclude Yukawa unification assumptions.

The parameter $\tan\beta$ may also be constrained by other processes.
For example, for low $\tan\beta$, chargino production at a linear
$e^+e^-$ collider may provide stringent constraints on
$\tan\beta$\cite{FMPT}.  This requires a sufficient Higgsino component
in the chargino, and the bound deteriorates for moderate and high
values of $\tan\beta$.  For high $\tan\beta$, there are a number of
possible probes.  If staus are pair produced at a future linear
collider, $\tilde{\tau}_L$--$\tilde{\tau}_R$ mixing\cite{Nojiri} may
be able to measure $\tan\beta$ sensitively in a range determined by
the Higgsino component of the lightest neutralino.  Alternatively, the
magnetic dipole moment of the muon, $(g-2)_{\mu}$, may be sensitive to
$\tan\beta \agt 20$ for slepton masses $m_{\tilde{l}} \alt 300$
GeV\cite{takeo}.  In addition, the discovery of $H,A \to
\tau\bar{\tau}$ at the LHC may be used to set the lower bound
$\tan\beta \agt 10$\cite{LHC}.  In general, however, the heavy Higgs
sector appears to be one of the most challenging for the LHC, and the
discovery and study of heavy Higgs bosons there may be
difficult\cite{GSW}. Of course, in the ideal case that measurements
confirm a particularly simple model, for example, the so-called
minimal supergravity model, in which all supersymmetric particle
masses and interactions are determined by only five additional
parameters, studies have shown that highly accurate measurements of
$\tan\beta$ may be obtained both at the LHC\cite{H} and
NLC\cite{SUSYNLC}.

It is interesting to note, however, that the power of these other
methods is usually greatest for low or high values of $\tan\beta$. In
addition, these measurements all involve many other SUSY parameters
and require certain conditions to be applicable.  In contrast, the
heavy Higgs measurement is most sensitive in the range where these
other measurements are weak, and is relatively free of other
assumptions.  It is clear, however, that no one process is powerful
throughout the range of $\tan\beta$ and for all scenarios.  Of course,
if more than one test is available, their consistency will be an
important test of SUSY.

\acknowledgements

We would like to thank J.~Bagger, A.~Djouadi, S.~Dong, K.~Fujii,
H.~Haber, I.~Hinchliffe, D.~Jackson, T.~Kon, H.~Murayama, M.~Nojiri,
and Y.~Yamada for valuable discussions and useful comments.  The
authors thank the organizers of Snowmass '96, and JLF gratefully
acknowledges the support of a JSPS Postdoctoral Fellowship and thanks
the KEK, CERN, and Rutgers Theory Groups for hospitality during the
course of this work.  This work was supported in part by the Director,
Office of Energy Research, Office of High Energy and Nuclear Physics,
Division of High Energy Physics of the U.S.  Department of Energy
under Contract DE--AC03--76SF00098 and in part by the NSF under grant
PHY--95--14797.

\appendix
\section*{Decay widths}

In this appendix we give formulas for the decay widths of the heavy
Higgs scalars to quarks and leptons for reference.  Additional decay
modes are $H^- \to W^- h$, $H \to W^+W^-, ZZ, hh, AA$, and $A\to Zh$,
as well as SUSY decay modes involving squarks, sleptons, charginos and
neutralinos.  All of these decay widths may be found in Appendix B of
Ref.~\cite{HHG}.

The charged Higgs boson decay width for fermion pairs $\bar{f}_u f_d$
is

\begin{equation}
\label{bra}
\Gamma ( H^- \to \bar{f}_u f_d ) = 
\frac{N_c g^2}{32 \pi m_W^2} \left( 
m_{f_d}^2 \tan^2\beta + m_{f_u}^2 \cot^2 \beta \right) 
m_{H^{\pm}} \left(1-\frac{m_{f_u}^2} {m_{H^{\pm}}^2} \right)^2 \ ,
\end{equation}
where $N_c$ is the number of color, and we have approximated $m_{f_d}
\ll m_{H^{\pm}}$ in the phase space factor.

For $A$ and $H$ decays to $f \bar{f}$, the width is given by 

\begin{equation}
\label{brb}
\Gamma ( A,H \to f \bar{f} ) = 
\frac{N_c g^2}{32 \pi m_W^2} 
m_{f}^2 C 
m_{A,H} \left(1-\frac{4m_{f}^2}{m_{A,H}^2} \right)^p \ ,
\end{equation}
where the coefficient $C$ and exponent $p$ are specified as follows:

\begin{eqnarray}
A \to f_u\bar{f}_u \em : \qquad & C=\cot^2 \beta, & \qquad p=1/2 \ ,\\
A \to f_d\bar{f}_d \em : \qquad & C=\tan^2 \beta, & \qquad p=1/2\ ,\\
H \to f_u\bar{f}_u \em : \qquad & C=\frac{\sin^2 \alpha}{\sin^2\beta}, &
\qquad p=3/2 \ ,\\ 
H \to f_d\bar{f}_d \em : \qquad & C=\frac{\cos^2 \alpha}{\cos^2\beta}, &
\qquad p= 3/2 \ .
\end{eqnarray}

\begin{table}
\caption{The cut parameters for our simulation.}
\begin{tabular}{l c c}
{} & {$\sqrt{s}=500$ GeV} & {$\sqrt{s}=1$ TeV} \\
\hline
{$\Delta E_{H^\pm}^{-}$}  & {30 GeV} & {30 GeV} \\
{$\Delta E_{H^\pm}^{+}$}  & {5 GeV}  & {10 GeV} \\
{$\Delta E_t^{\rm 1b}$}   & {15 GeV} & {25 GeV} \\
{$\Delta E_t^{\rm 2b}$}   & {20 GeV} & {35 GeV} \\
{$\Delta m_Z^{-}$}        & {30 GeV} & {40 GeV} \\
{$\Delta m_Z^{+}$}        & {5 GeV}  & {10 GeV} \\
{$\Delta m_W$}            & {10 GeV} & {15 GeV} \\
\end{tabular}
\label{table:cut_params}
\end{table}

\begin{table}
\caption{The standard model background for channels 1 -- 8 (in fb).
Here, we assume $\epsilon_b=60\%$ and $\epsilon_c=2.6\%$.}
\begin{tabular}{c c c c c c c c c}
{} & {1} & {2} & {3} & {4} & {5} & {6} & {7} & {8}\\
\hline
{$\protect\sqrt{s}=500$ GeV} &
{0.11} & {0.071} & {1.5} & {6.2} & 
{1.2} & {0.12} & {0.17} & {0.0053} \\
{$\protect\sqrt{s}=1$ TeV} &
{0.36} & {0.35} & {1.0} & {3.0} & 
{0.47} & {0.22} & {0.30} & {0.0095} \\
\end{tabular}
\label{table:bg}
\end{table}

\begin{table}
\caption{Cumulative efficiency of cuts 1a -- 1e for $\protect
\sqrt{s}=500$ GeV. Here, we take $m_{H^{\pm}} = 200$ GeV and
$\tan\beta = 5$. The $b$-tagging efficiency is not included.}
\begin{tabular}{c c c c}
{cut} & {$H^+ H^-$} & {$t \bar{t}$} & {$A H$} \\
\hline
{1a} & {$0.90$} & {$1.2\times 10^{-3}$} & {$4.5\times 10^{-3}$} \\
{1b} & {$0.82$} & {$1.1\times 10^{-3}$} & {$4.5\times 10^{-3}$} \\
{1c} & {$0.51$} & {$5.7\times 10^{-4}$} & {$3.8\times 10^{-3}$} \\
{1d} & {$0.51$} & {$5.7\times 10^{-4}$} & {$3.8\times 10^{-3}$} \\
{1e} & {$0.45$} & {$5.6\times 10^{-4}$} & {$3.4\times 10^{-3}$} \\
\end{tabular}
\label{table:1}
\end{table}

\begin{table}
\caption{Cumulative efficiency of cuts 2a -- 2e for $\protect
\sqrt{s}=500$ GeV. Here, we take $m_{H^{\pm}} = 200$ GeV and
$\tan\beta = 60$.  The $tbH^{\pm}$ signal is normalized to 1 after cut
2a.  The $b$-tagging efficiency is not included.}
\begin{tabular}{c c c c}
{cut} & {$tbH^{\pm}$} & {$t \bar{t}$} & {$A H$} \\
\hline
{2a} & {$1.00$} & {$0.95$}              & {$0.33$} \\
{2b} & {$0.65$} & {$3.1\times 10^{-4}$} & {$0.27$} \\
{2c} & {$0.38$} & {$2.2\times 10^{-4}$} & {$0.27$} \\
{2d} & {$0.38$} & {$2.1\times 10^{-4}$} & {$0.26$} \\
{2e} & {$0.33$} & {$1.5\times 10^{-4}$} & {$0.24$} \\
\end{tabular}
\label{table:2}
\end{table}

\input psfig

\noindent

\noindent

\noindent

\noindent

\noindent

\noindent

\noindent

\noindent

\noindent

\noindent

\noindent

\noindent

\noindent

\noindent

\noindent

\noindent

\noindent

\noindent

\noindent

\noindent

\noindent

\end{document}